\documentclass[12pt]{article}
\usepackage[utf8]{inputenc}
\usepackage[pdftex]{graphicx}
\usepackage[a4paper ,margin=1.375in]{geometry} %
\usepackage{authblk}
\usepackage{amsmath}
\usepackage{amssymb}
\usepackage{natbib}
\usepackage{tikz}
\usepackage[colorlinks,citecolor=blue]{hyperref}
\usepackage{pifont}
\usepackage{lineno}
\usepackage{setspace}
\newcommand{\cmark}{\ding{51}}%
\newcommand{\xmark}{\ding{55}}%

\title{Desiderata for normative models of synaptic plasticity}
\date{}
\providecommand{\keywords}[1]
{
  \small	
  \textbf{Keywords: } #1
}

\author[1,2, $\dag$]{\large Colin Bredenberg}
\author[1,3]{\large Cristina Savin}
\affil[1]{ \normalsize Center for Neural Science,
           New York University, New York, NY 10003, USA}
\affil[2]{ \normalsize Mila- Quebec AI Institute, 6666 Rue Saint-Urbain, Montréal, QC H2S 3H1}
\affil[3]{ \normalsize Center for Data Science,
           New York University, New York, NY 10011, USA}
\affil[$\dag$]{\normalsize Corresponding author: colin.bredenberg@mila.quebec}

\newcommand{\comment}[1]{}

\newcommand{\derivative}[2]{\frac{d#1}{d#2}}
\newcommand{\weight}{\mathbf{W}}
\newcommand{\win}{\mathbf{W}^{in}}

\newcommand{\rate}{\mathbf{r}}
\newcommand{\stim}{\mathbf{s}}

\newcommand{\decoder}{\mathbf{W}^{out}}
\newcommand{\out}{o}
\newcommand{\target}{\hat o}

\newcommand{\expect}[1]{\mathbb{E} \left [ #1 \right ]}

\newcommand{\noise}{\boldsymbol{\eta}}
\newcommand{\reward}{R(\rate, \stim)}
\newcommand{\obj}{\mathcal{O}(\win)}

\begin{document}
\maketitle
\keywords{computational neuroscience, learning, synaptic plasticity}
\begin{abstract}\noindent
Normative models of synaptic plasticity use a combination of mathematics and computational simulations to arrive at predictions of behavioral and network-level adaptive phenomena. In recent years, there has been an explosion of theoretical work on these models, but experimental confirmation is relatively limited. In this review, we organize work on normative plasticity models in terms of a set of desiderata which, when satisfied, are  designed to guarantee that a model has a clear link between plasticity and adaptive behavior, consistency with known biological evidence about neural plasticity, and specific testable predictions. We then discuss how new models have begun to improve on these criteria and suggest avenues for further development. As prototypes, we provide detailed analyses of two specific models -- REINFORCE and the Wake-Sleep algorithm. We provide a conceptual guide to help develop neural learning theories that are precise, powerful, and experimentally testable.
\end{abstract}

\section{Introduction}
Our identities change with time, gradually reshaping our experiences. We remember, we associate, we learn. However, we are only beginning to understand how changes in our minds arise from underlying changes in our brains.
Of the many features of neural architecture that are altered over time, from the biophysical properties of individual neurons to the creating or pruning of synapses between neurons, changes in the strength of existing synapses have long been among the most prominent candidates for the neural substrate of longitudinal perceptual and behavioral change, because many synaptic connections are easily modified, and these modifications can persist for extended periods of time \citep{bliss1993synaptic}. Further, synaptic modification has been associated with many of the brain's critical adaptive functions, including memory \citep{martin2000synaptic}, experience-based sensory development \citep{levelt2012critical}, operant conditioning \citep{ohl2005learning, fritz2003rapid}, and compensation for stroke \citep{murphy2009plasticity} or neurodegeneration \citep{zigmond1990compensations}.
However, beyond these associations, a {precise link} between plasticity and adaptive behaviors of interest is currently lacking.

Here, we distinguish `normative' modeling approaches from other alternatives, demonstrate why they show promise for establishing this link, and outline a set of desiderata which articulate how recent progress on normative plasticity models strengthens the link between plasticity and system-wide adaptive phenomena. To provide concrete examples of these principles in action, in Appendices \ref{reinforce_tutorial} and \ref{wake_sleep_tutorial} we provide worked tutorials on two complementary canonical normative plasticity models---REINFORCE \citep{williams1992simple} for reinforcement learning, and the Wake-Sleep algorithm for unsupervised learning \citep{dayan1995helmholtz, hinton1995wake}---and illustrate their successes and failures to match our desiderata.

\subsection{Phenomenological, mechanistic, and normative plasticity models}
\begin{figure}[t!]
	\includegraphics[width=\linewidth]{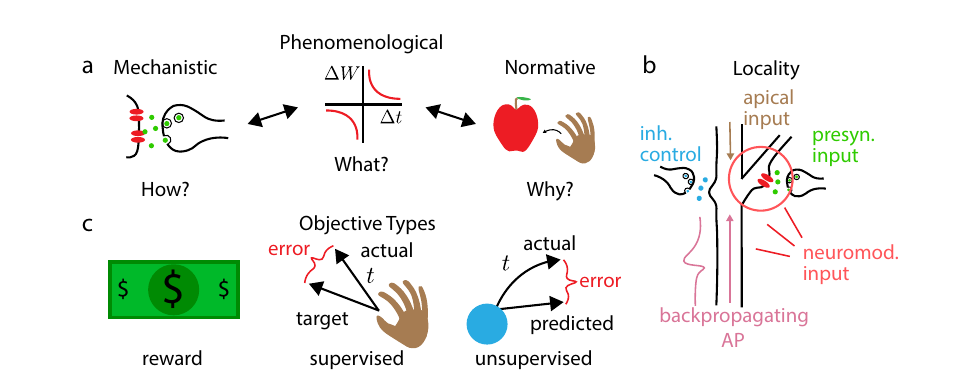}
    \centering
	\caption{\small {\bf Defining normative modeling.} {\bf a.} Spectrum of synaptic plasticity models. Mechanistic models show how detailed biophysical interactions produce observed plasticity, phenomenological models concisely describe what changes in experimental variables (e.g. post-pre relative spike timing $\Delta t$) affect plasticity ($\Delta W$), and normative models explain why the observed plasticity implements capabilities that are useful to the organism. {\bf b.} Schematic illustrating the range of local variables that may be available for synaptic plasticity. These include, but are not limited to: backpropagating action potentials from the soma, apical dendritic input, pre- and postsynaptic activity, neuromodulatory signals, and potentially inhibitory input from local microcircuitry. {\bf c.} Classes of objective function used in normative plasticity theories. Reward-based objectives involve only feedback about how well the organism or network performed, whereas supervised objectives provide explicit targets for network output. By contrast, unsupervised objectives do not require any form of explicit feedback to train the network.}
	\label{fig_1}
\end{figure}

We distinguish between three partially overlapping types of model: phenomenological, mechanistic, and normative (Fig. \ref{fig_1}a) \citep{levenstein2020role}. The focus of this review is normative plasticity models, but to understand their importance, we first describe their relationship to their counterparts.

In the simplest terms, a phenomenological model's focus is on describing experimental data: the primary goal is to concisely summarize relationships between observed variables. As an example, many early studies of spike-timing-dependent plasticity (STDP) described the relationship between plasticity and the relative timing of pre- and post-synaptic spikes with exponential curves fit to data \citep{zhang1998critical, dan2004spike, sjostrom2010spike}. Such models can reduce the complexity of data, providing interpretability and, to some extent, predictive power. They are incomplete descriptions of the biophysical processes that form the causal link between spike times and plasticity, but extract and summarize important features of the data on which subsequent theories and models can build.
    
A mechanistic model attempts to explain a set of experimental results in terms of causal interactions between biophysical quantities. For instance, since the initial characterization of STDP, a plethora of studies have emerged characterizing in detail the interactions between backpropagating action potentials \citep{magee1997synaptically}, dendritic morphological properties \citep{froemke2005spike, letzkus2006learning, sjostrom2006cooperative}, local membrane voltage, NMDA ion channel properties, and calcium-sensitive molecules near the synapse. Mechanistic models \citep{graupner2010mechanisms} characterize how these variables all collectively contribute to the strengthening or weakening of the synapse. As a consequence of their depth and breadth, mechanistic models can often provide predictions that are outside of the scope of the original experiment, and provide useful targets for experimental manipulation.

        
The distinction between phenomenological models and mechanistic models is not always completely crisp, especially in areas where our scientific understanding is progressing rapidly. In nascent mechanistic models, there often exist `black boxes' that specify interactions between known biophysical quantities, without a precise understanding of whether or how these interactions are implemented \citep{craver2007explaining}. Because they lack a direct relation to well-understood biophysics, these `black boxes' act in essentially the same way as variables do in a phenomenological model. In this way, we can see that there exists a spectrum between phenomenological and mechanistic models, and that oftentimes, mechanistic models grow from phenomenological ones. However, there is more to the spectrum: while phenomenological and mechanistic models articulate how synaptic plasticity works, they do not explain \textit{why} it exists in the brain, i.e. what its importance is for neural circuits, behavior, or perception. To answer this question with any precision requires an appeal to normative modeling.

Normative models aim answer this `why' question by connecting plasticity to observed network-level or behavioral-level phenomena, including memory formation \citep{hopfield1982neural} and consolidation \citep{benna2016computational, clopath2008tag, fusi2005cascade}, reinforcement learning \citep{fremaux2016neuromodulated}, and representation learning \citep{hinton1995wake, oja1982simplified, rao1999predictive, savin2010independent}.
This class of plasticity model, in our view, employs a fundamentally different set of methodologies from phenomenological or mechanistic models, in order to provide the missing link between plasticity and function. Guided by the intuition that plasticity processes have developed on an evolutionary timescale to near-optimally perform adaptive functions, normative plasticity theories are typically `top-down', in that they begin with a set of prescriptions about how synapses `should' modify in order to optimally perform a given learning-based function. Subsequently, with varying degrees of success, these theories attempt to show that real biology matches or approximates this optimal solution.
As an example, an increasing body of literature is establishing a correspondence between classical reinforcement learning algorithms \citep{williams1992simple} and reward-modulated Hebbian synaptic plasticity models of learning in the brain \citep{fremaux2016neuromodulated}. This process is ongoing, and though experimental support for such forms of plasticity are growing \citep{gerstner2018eligibility}, much work remains to be done. Similar efforts are underway to construct approximations to the backpropagation algorithm which can serve as models of neural plasticity \citep{marschall2020unified, lillicrap2020backpropagation, richards2019dendritic, urbanczik2014learning}.
Here, we will review classical normative plasticity approaches and discuss recent efforts to improve upon them. 

\section{Desiderata for normative models}
One of the biggest challenges for a normative model of synaptic plasticity is its connection to biology: artificial neural networks with simulated synapses (synaptic weight parameters) that adapt to improve performance on any of a variety of functions from sensory processing \citep{lecun1989handwritten, krizhevsky2012imagenet}, to motor learning \citep{heess2017emergence, hafner2019dream}, to abstract game learning \citep{silver2017mastering, vinyals2019grandmaster} are much more accessible to mathematical and empirical investigation than the neural circuits implementing these functions in the brain. Compared to the simulations and mathematical analysis used to explore machine learning algorithms, neuroscience experiments are time-consuming and expensive. Further, network simulations provide total access to neural activations, stimuli, and synaptic parameters over the whole course of learning, whereas any one neuroscience experiment can only reveal a very small amount about what is going on in a circuit. Therefore, it is a major challenge to identify how to improve normative models with relatively limited access to experimental data confirming or rejecting their predictions.

In what follows, we will articulate a set of desiderata that can serve as both an organizing tool for understanding the contributions of recent normative plasticity modeling efforts and as intermediate objectives for the development of new models in the absence of explicit experimental rejection or confirmation of older work.
We will argue that each principle is desirable for some combination of the following reasons: first, it may help ensure that the plasticity model actually qualifies as normative; second, it may require a model to accommodate known facts about biology; third, it may ensure that models can be compared properly to existing experimental literature and generate genuinely testable experimental predictions.
Most of these desiderata are relatively intuitive and simple. However, it has proven incredibly difficult for existing models of any adaptive cognitive phenomenon---from sensory representation learning, to associative memory formation, to reinforcement learning---to satisfy all desiderata in tandem.

\subsection{Improving performance} \label{objective}
One way to view the normative approach is that it attempts to organize the diversity of synaptic dynamics existing within a neural system into the simplest explanatory framework possible for what functions the system's plasticity subserves. Usually, this framework is mathematical for pragmatic reasons: mathematics provides the precision and power necessary to establish clear relationships between plasticity and function. In particular, viewing neural plasticity as an approximate optimization process has been very fruitful \citep{lillicrap2020backpropagation, richards2019deep}, wherein synaptic modifications progressively reduce a scalar loss function. This process can be divided into two steps: articulating an appropriate objective, and subsequently demonstrating that a synaptic plasticity mechanism improves performance on that objective.

It can be extremely difficult to reduce the full range of functions a given circuit must perform to a scalar objective function, but as we will show subsequently, the conceptual benefits can be immense. On one side, picking too simple an objective function runs the risk of ignoring many functions a system is required to perform. For instance, early normative theories of learning in sensory systems show how synaptic plasticity could minimize the objective function underlying principal component analysis (PCA) \citep{oja1982simplified}, but merely representing the principal components of an incoming sensory stream is an inadequate characterization of sensory processing for several reasons. PCA can capture only second-order properties (mean and covariance) of naturalistic stimuli and does not perform the highly nonlinear processing required for cortical neurons to exhibit gain control capabilities \citep{simoncelli1998model} and texture \citep{ziemba2016selectivity} and object class \citep{rust2010selectivity} selective responses. A given synaptic plasticity mechanism may only be able to minimize a restricted subset of objectives, and for a normative theory, the set of possible objectives that can be minimized must encompass a wide range of functions that the brain is known to subserve. Beyond principal component analysis, many modern models of unsupervised representation learning use objectives for training hierarchical generative models (e.g. the evidence lower bound (ELBO) which underlies the Wake-Sleep algorithm \citep{dayan1995helmholtz} and predictive coding \citep{rao1999predictive}, and allows for multilayer, nonlinear representation learning). On the other side, selecting too flexible an objective function can run the risk of `overfitting' experimental data, a problem that is particularly salient for Bayes-optimal accounts of neuroscientific and psychological phenomena \citep{bowers2012bayesian}. As an extreme example, if we were to postulate that the `objective' of a neural system is to behave exactly as it is observed to behave experimentally, i.e. everything in a neural system happens precisely as was `intended', then the normative project becomes vacuous: the model provides neither conceptual simplification nor predictive power beyond what was observed experimentally, and has consequently failed to provide a useful explanation of the data. Therefore, the quality of an objective function is determined by both how many phenomena it is able to explain and how simple it is.

%

    
Normative theories of synaptic plasticity developed to date usually involve some combination of supervised, unsupervised, or reinforcement learning objectives (Fig. \ref{fig_1}c). The choice of objective function for a neural system is laden with philosophical assumptions about the system's functional utility, and can exert a huge influence on the resultant form and scope of applicability of the synaptic plasticity model. For instance, supervised learning usually involves the existence of either an internal or external teacher. If the teacher is external, such a learning mechanism could only be leveraged under the very specific and comparatively rare conditions in which the organism is being overtly taught, as is the case, for instance, in some models of zebra finch song learning \citep{fiete2007model}. If the teacher is internal, a plausible normative theory is limited in the types of knowledge the `self-supervisor' may reasonably construct and provide (for instance, motor error signals \citep{gao2012distributed, bouvier2018cerebellar} or saccade information indicating that a visual scene has changed \citep{illing2021local}).
Generative modeling is a form of unsupervised learning that postulates that a sensory system is actively building a probabilistic model of its sensory inputs, which can be used to simulate possible future outcomes and perform Bayesian reasoning \citep{fiser2010statistically}. This vision of sensory coding is popular both for its ability to accommodate normative plasticity theories \citep{rao1999predictive, dayan1995helmholtz,kappel2014stdp, bredenberg2021impression} and for its philosophical vision of sensory processing as a form of advanced model building, beyond simple sensory transformations. However, model construction is only indirectly useful for many tasks involving rewards and planning, and so such plasticity would have to occur concomitantly with reward-based \citep{fremaux2016neuromodulated} or motor \citep{gao2012distributed, feulner2021neural} learning. Furthermore, alternative perspectives on sensory processing exist, including those based on maximizing the information about a sensory stimulus contained in a neural population \citep{attneave1954some, atick1990towards} subject to metabolic efficiency constraints \citep{tishby2000information, simoncelli2001natural}, and those based on `contrastive methods' \citep{oord2018representation, illing2021local}, where a self-supervising internal teacher encourages the neural representation of some stimuli to grow closer together, while encouraging others to grow more discriminable.

Evaluating which objective function (or functions) best explains the properties of a neural system is very hard: while some forms of objective function may have discriminable effects on plasticity (e.g. supervised vs. unsupervised learning \citep{nayebi2020identifying}), others are even provably impossible to distinguish. As a simple example, suppose that we have an $N^r$ dimensional single-layer neural network receiving $N^s$ dimensional stimuli through an $N^r \times N^s$ dimensional weight matrix $\weight$. We have the response given by:
\begin{equation}
    \rate = f(\weight \stim),
\end{equation} 
where $f(\cdot)$ is a $\tanh$ nonlinearity. Now suppose that some setting of synaptic weights $\weight^*$ minimizes an objective function $\mathcal{L}$, i.e. $\mathcal{L}(\weight^*) \leq \mathcal{L}(\weight) ~ \forall \weight$. We might be tempted to argue that because $\weight^*$ minimizes $\mathcal{L}$, $\mathcal L$ must be the objective that the system is minimizing. However, there are an infinite variety of alternative objectives that share this same minimum (Appendix \ref{identifiability}). This motivates the idea that for a given dataset, it is very plausible that one objective ($\tilde{\mathcal{L}}$) can \textit{masquerade} as another ($\mathcal{L}$). In some cases, complex objective functions can masquerade as simple objectives, which may only be epiphenomenal. For instance, it has been hypothesized that synaptic modifications may preserve the balance between inhibitory and excitatory inputs to a cell \citep{vogels2011inhibitory}; recent theories have proposed that this E/I balance may only be a consequence of a more advanced theory of sensory predictive coding \citep{brendel2020learning}. In other cases, philosophically distinct frameworks, such as generative modeling, information maximization, or denoising may simply produce similar synaptic plasticity modifications because the frameworks often overlap heavily \citep{vincent2010stacked}, and may not be distinguishable on simple datasets without targeted experimental attempts to disambiguate between the two perspectives.
    
Furthermore, not every function performed by biological systems has been adequately incorporated into a simple optimization framework. For example, though the Hebbian plasticity rule used in Hopfield networks endows model circuits with associative memory, the utility of learning is characterized by the dynamical attractor structure it embeds in the neural circuit, rather than by its direct minimization of an objective function \citep{hopfield1982neural}. In addition, the notion that some parts of the brain may have synaptic plasticity mechanisms for representation learning while other parts have plasticity for reinforcement learning suggests that the brain may be better viewed as a collection of interacting systems with only partially overlapping goals. This multiagent \citep{zhang2021multi} formulation of learning has intuitive appeal, because it can decompose broad objectives like survival into a series of intermediate objectives carried out by individual systems. Such a formulation could help explain how locality emerges, i.e. why synapses do not need information about distant neural circuits in order to improve performance. However, with this additional appeal comes additional conceptual and mathematical complexity, because improving performance on one objective could very easily harm the performance of other systems. Therefore, insofar as a collection of neural circuits and plasticity mechanisms \textit{can} be viewed as acting in concert to improve a unified objective, simple optimization is the preferable perspective.

Having addressed many difficulties associated with choosing a good objective function, we now move to difficulties involved in demonstrating that a particular synaptic plasticity rule decreases a chosen objective\footnote{Some objectives (like reward functions) are best thought of as being maximized rather than minimized. Without loss of generality, in such cases we can minimize the negative reward function.}. How could such a property be proven? For a particular plasticity rule to reduce an objective, we need to show that the following principle holds:

\begin{align} 
    &\mathcal{L}(\weight + \Delta \weight) < \mathcal{L}(\weight) \nonumber \\
    \Rightarrow &\mathcal{L}(\weight + \Delta \weight) - \mathcal{L}(\weight) < 0,
\end{align}
for some update $\Delta \weight$ determined by the plasticity rule. If we accept the additional supposition that $\Delta \weight$ is very small, we can employ the first order Taylor approximation (treating $\weight$ as a flattened vector of length $N^r \times N^s$): $\mathcal{L}(\weight + \Delta \weight) \approx \mathcal{L}(\weight) + \derivative{\mathcal{L}}{\weight} (\weight)^T \Delta \weight$. Substituting this approximation into our reduction criterion, we have after cancellation:
\begin{equation} \label{inner_product}
    \derivative{\mathcal{L}}{\weight}(\weight)^T \Delta \weight < 0.
\end{equation}
This shows that for small weight updates (slow learning rates), the inner product between a synaptic learning rule $\Delta \weight$ and the gradient of the selected loss function $\mathcal{L}(\weight)$ with respect to the weight change must be negative. The simplest way to ensure that this is true is for $\Delta \weight$ to equal a small scalar $\lambda$ times the negative gradient of the loss ($- \lambda \derivative{\mathcal{L}}{\weight}(\weight)^T \derivative{\mathcal{L}}{\weight} (\weight) = - \lambda \| \derivative{\mathcal{L}}{\weight}(\weight)\|_2^2 < 0$). If this were true, plasticity would be guaranteed to improve performance on the objective $\mathcal{L}$. Unfortunately, for even the simplest neural networks and objective functions,  naive methods of calculating this gradient will prove to be nonlocal (see Appendix \ref{weight_transport} for a simple example). Thus, the critical challenge for normative theories of synaptic plasticity is finding ways that neural networks can find synaptic modifications $\Delta \weight$ that demonstrably have a negative inner product with the gradient of a desired objective $\mathcal{L}$, while still allowing the neural network to satisfy biologically realistic locality constraints. However, it is important to note that if an update $\Delta \weight$ reduces any one objective function, then it also reduces an infinite number of similar alternative objective functions (Appendix \ref{identifiability}); therefore it is perhaps best to think of normative plasticity models in terms of the family of objective functions that they minimize---committing to any one particular objective within that family reflects the predilections of the theorist, not the system.
    
Different normative studies demonstrate that Eq. \ref{inner_product} holds by different methods. Some studies show empirically across many simulations that this inner product is negative \citep{lillicrap2016random, marschall2020unified}. However, these demonstrations alone do not answer the following questions: how would we know that the network would still perform well if a different task were chosen, or if the network's architecture were different, or if various elements of the simulated plasticity mechanism were changed? A simulation has relatively limited power to extrapolate beyond its immediate results, especially when the neuron models used in large-scale network simulations are often very reductive \citep{gerstner2002spiking} and when small changes in simulated network parameters can effect large qualitative differences in network behavior \citep{xiao2021data}.
Further, a battery of \textit{in silico} simulations under a variety of different parameter settings and circumstances rapidly begins to suffer the curse of dimensionality, becoming almost as extensive as the collection of \textit{in vivo} or \textit{in vitro} experiments that it is attempting to explain. As such, simulation-based justifications suffer from a lack of conciseness and an inability to easily address counterfactuals.

For this reason, much focus in the field has been devoted to constructing mathematical arguments as to why Eq. \ref{inner_product} should hold for a given local synaptic plasticity rule. Some plasticity rules amount to stochastic approximations to the true gradient \citep{williams1989learning, scellier2017equilibrium} and some are systematically biased but maintain a negative inner product under reasonable assumptions \citep{bredenberg2021impression, dayan1995helmholtz, amari1999convergence, meulemans2020theoretical}. Mathematical analysis allows one to know quite clearly when a particular plasticity rule will decrease a loss function, and identifies how plasticity mechanisms should change with changes in the network architecture or environment. However, analysis is often only possible under restrictive circumstances, and it is often necessary to supplement mathematical results with empirical simulations in order to demonstrate that the results extend to more general, more realistic circumstances.

\subsection{Locality} \label{locality}
Biological synapses can only change strengths using chemical and electrical signals available at the synapse itself. `Locality' refers to the idea that a postulated synaptic plasticity mechanism should only refer to variables that could be conceivably available at a given synapse (Fig. \ref{fig_1}b). Though locality may seem like an obvious requirement for any theory of biological function, for synaptic plasticity it presents a great mystery: how does a system as a whole, whose success or failure is determined by the joint action of many neurons distributed across the entire brain, communicate information to individual synapses about how to improve? The success of most machine learning algorithms relies on nonlocal, even global, propagation of learning signals, including backpropagation \citep{werbos1974beyond, rumelhart1985learning} (See Appendix \ref{weight_transport}), backpropagation through time \citep{werbos1990backpropagation}, and real-time recurrent learning \citep{williams1989learning}.

Despite its importance as a guiding principle for normative theories of synaptic plasticity, locality is a slippery concept, primarily because of our insufficient understanding of the precise battery of biochemical signals available to a synapse, and how those signals could be used to approximate quantities required by theories.
As a simple example, many normative theories require information about the pre- and postsynaptic firing rates of a neuron, similar to Hebb's Postulate \citep{hebb1949organisation}. However, neurons predominately communicate to one another through discrete action potentials, and additional cellular machinery would be required to form an estimate of pre- and postsynaptical firing rates based on backpropagating action potentials from the soma and on postsynaptic potentials. Whether a plasticity rule derived from normative principles involves rate or spike-based information is often a function of the neuron model used in the theory, and it is often difficult to formulate predictions about how a realistic, non-idealized neuron should exactly modify its synapses based on over-simplified models. Therefore,  normative theories typically declare success when some standard of plausibility is reached, where derived plasticity rules roughly match the experimental literature \citep{payeur2021burst} or only require reasonably simple functions of postsynaptic and pre-synaptic activity that a synapse could hypothetically approximate \citep{oja1982simplified, scellier2017equilibrium, williams1992simple}.

In normative models of synaptic plasticity, the requirement of locality is in perpetual tension with the general requirement for some form of `credit assignment' \citep{lillicrap2020backpropagation, richards2019deep}, i.e. a mechanism capable of signaling to a neuron that it is `responsible' for a network-wide error, and should modify its synapses to reduce errors. Depending on a network's objective, a system's credit assignment mechanism \textit{could} take a wide variety of forms, some small number of which may only require information about the pre- and post-synaptic activity of a cell \citep{oja1982simplified, pehlevan2015hebbian, pehlevan2017similarity, obeid2019structured, brendel2020learning}, but many of which appear to require the existence of some form of error \citep{scellier2017equilibrium, lillicrap2016random, akrout2019using} or reward-based \citep{williams1992simple, fiete2007model, legenstein2010reward} signal.

The extent to which a credit assignment signal postulated by a normative theory meets the standards of `locality' depends heavily on the nature of the signal. For instance, there is growing support for the idea that neuromodulatory systems, distributing dopamine \citep{otani2003dopaminergic, calabresi2007dopamine, reynolds2002dopamine}, norepinephrine \citep{martins2015coordinated}, oxytocin \citep{marlin2015oxytocin}, and acetylcholine \citep{froemke2013long, guo2019cholinergic, hangya2015central, rasmusson2000role, shinoe2005modulation} signals can propagate information about reward \citep{guo2019cholinergic}, expectation of reward \citep{schultz1997neural}, and salience \citep{hangya2015central} diffusely throughout the brain to induce or modify synaptic plasticity in their targeted circuits. Therefore, it may be reasonable for normative theories to postulate that synapses have access to global reward or reward-like signals, without violating the requirement that plasticity be affected only by locally-available information \citep{fremaux2016neuromodulated}.

Locality as a desideratum serves as a heuristic stand-in for the requirement that a normative model must be eventually held to the standard of experimental evidence. This is not to say that normative models cannot postulate neural mechanisms that have not yet been observed experimentally. However, for such an exercise to be constructive, the theory should clearly articulate how it deviates from the current state of the experimental field, and how these deviations can be tested (Section \ref{testability}; see Appendices \ref{reinforce_tutorial} and \ref{wake_sleep_tutorial} for concrete examples of this process). Furthermore, the process of mathematical abstraction necessitates approximation \citep{cartwright1984laws}: constraining a normative theory to adhere to `locality' without necessarily requiring a perfect correspondence to experimental data allows normative theories to strive to capture the essence of synaptic learning processes without becoming mired in technical details.

\subsection{Architectural plausibility} \label{architecture}
\begin{figure}[t!]
	\includegraphics[width=\linewidth]{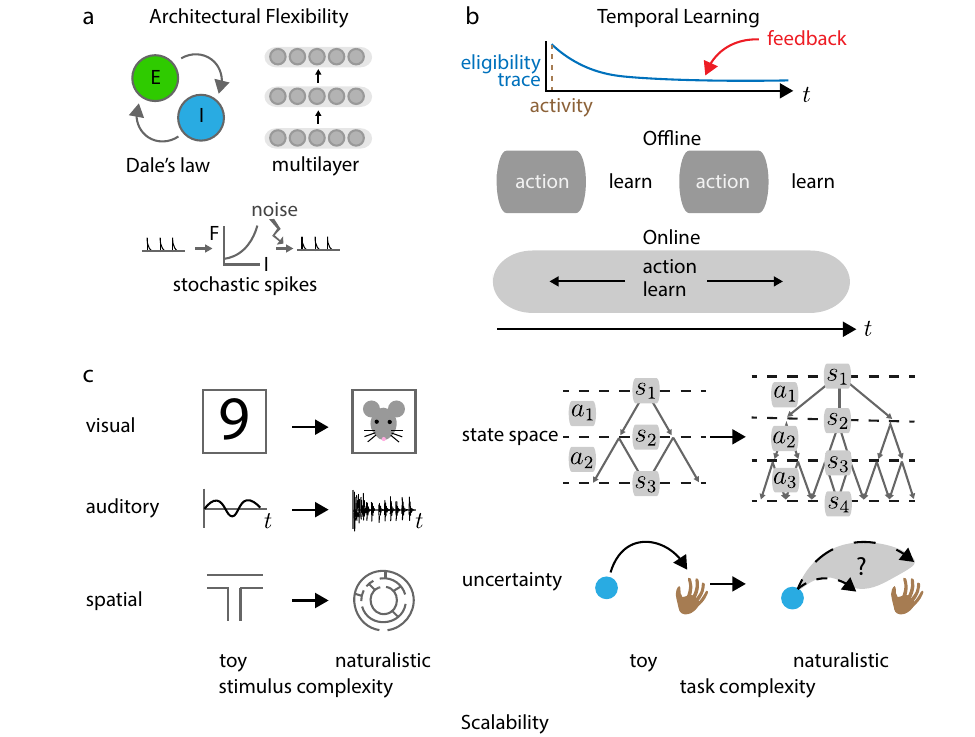}
    \centering
	\caption{ \small {\bf Architecture and scalability considerations for normative plasticity models.} {\bf a.} Features of realistic biological networks that normative plasticity theories should be able to account for: separation of excitatory and inhibitory neuron populations; stochastic and spiking input-output functions for individual neurons; and multilayer, recurrent connectivity. {\bf b.} For actions in the past to be associated with delayed supervisory or reinforcement signals, plasticity algorithms require a mechanism of temporal association. One candidate is the `eligibility trace,' which stores information about coactivity throughout time locally to a synapse, and subsequently modifies synaptic connections when paired with feedback information. Learning can occur offline, where some or all synaptic modification occurs in the absence of action or perception by the organism. Alternatively, it can occur online, where the organism acts and learns simultaneously. {\bf c.} Stimuli (left) and task structure (right) can become complex in many ways. Different sensory features (e.g. visual, auditory, or spatial information) can all be made more naturalistic by training networks on stimuli organisms are exposed to and learn from in natural environments. Further, tasks can be made more naturalistic by increasing the number of action options ($a$) and sequential state ($s$) transitions required for a network to achieve its goals and by adding uncertainty into the task.}
	\label{fig_2}
\end{figure}
The learning algorithm implemented by a plasticity model often requires specific architectural motifs to exist in a neural circuit in order to deliver reward, error, or prediction signals. These might include diffuse neuromodulatory projections (Fig. \ref{fig_supp_1}b) or neuron-specific top-down synapses onto apical dendrites (Fig. \ref{fig_supp_2}c). Such architectural features (or alternative, isomorphic motifs) are \textit{required} for the learning algorithm in question, and are known to exist in a wide range of cortical areas. However, normative plasticity models should not depend on circuit features that have been demonstrated not to exist in the modeled system, because spurious architectural features can be used to `cheat' at achieving locality by postulating unrealistic credit assignment mechanisms (see Appendix \ref{weight_transport}). Further, models lacking important features of neural circuits can be difficult to relate to experimental data.
In what follows, we will highlight several particularly important architectural motifs that have been the focus of recent work.

Contrary to the highly reduced deterministic rate-based models typically used in machine learning, neurons communicate through roughly discrete action potentials. Further, they exhibit numerous forms of variability due in part to synaptic failures and constant receipt of task-irrelevant signals (Fig. \ref{fig_2}a) \citep{faisal2008noise}. Normative theories which employ rate-based activations \citep{bredenberg2020learning, scellier2017equilibrium} or which assume that the input-output function of neurons is approximately linear \citep{oja1982simplified}, may not extend to the more realistic discrete, stochastic, and highly nonlinear setting. Further, by ignoring spike timing, such theories inherently produce plasticity rules that ignore the precise relationship between pre- and post-synaptic spike times, and will consequently be unable to capture STDP results. This both limits the expressive power of such models, and prevents their experimental validation. Fortunately, several methods which were originally formulated using rate-based models have subsequently been extended to spiking network models to great effect. Reward-based Hebbian plasticity based on the REINFORCE algorithm (Appendix \ref{reinforce_tutorial}) \citep{williams1992simple} has been generalized to stochastic spiking networks \citep{fremaux2013reinforcement}, while backpropagation approximations \citep{murray2019local} and predictive coding methods \citep{rao1999predictive} have subsequently extended to deterministic spiking networks \citep{bellec2020solution, brendel2020learning}. Therefore, a lack of a generalization to spiking networks is not necessarily a death knell for a normative theory, but many existing theories lack either an explicit generalization to spiking or a clear relationship to STDP, and the mathematical formalism that defines these methods may require significant modification to accommodate the change.

Real biological networks have a diversity of cell types with different neurotransmitters and connectivity motifs. At the bare minimum, a normative model must be able to accommodate Dale's Law (Fig. \ref{fig_2}a), which stipulates that the neurotransmitters released by a neuron are either excitatory or inhibitory, but not both (for the most part \citep{o198550th}). Though this might seem like a simple principle, enforcing Dale's principle can seriously damage the performance of artificial neural networks without careful architectural considerations \citep{cornford2021learning}. Furthermore, the mathematical results of \textit{many} canonical models of synaptic modification rely on symmetric connectivity between neurons, including Hopfield networks \citep{hopfield1982neural}, Boltzmann machines \citep{ackley1985learning}, contrastive Hebbian learning \citep{xie2003equivalence}, and predictive coding \citep{rao1999predictive}; this symmetry is partially related to the symmetric connectivity required by the backpropagation algorithm (Appendix \ref{weight_transport}). Symmetric connectivity means that the connection from neuron A to neuron B must be the same as the reciprocal connection from neuron B to neuron A. It inherently violates Dale's Law, because it means that entirely excitatory and entirely inhibitory neurons can never be connected to one another: the positive sign for one synapse and the negative sign for the reciprocal connection violates symmetry. Some models, such as Hopfield networks \citep{sompolinsky1986temporal} and equilibrium propagation \citep{ernoult2020equilibrium} have been extended to demonstrate that moderate deviations from symmetry can exist and still preserve function. Further, a recent mathematical reformulation of predictive coding has demonstrated that inter-layer symmetric connectivity is not necessary \citep{golkar2022constrained}. Therefore, recent results indicate that many canonical models believed to depend on symmetric connectivity can be improved upon.


Many early plasticity models, including Oja's rule \citep{oja1982simplified} and perceptron learning \citep{rosenblatt1958perceptron}, as well as more modern model recurrent network models focused on learning temporal tasks \citep{murray2019local} are designed to greedily optimize layer-wise objectives, and their mathematical justifications do not generalize to multi-layer architectures. Though greedy layer-wise optimization may be sufficient for some forms of unsupervised learning \citep{illing2021local}, a method that cannot account for how credit assignment signals are passed between cortical areas will not in general be able to support many complex supervised or reinforcement learning tasks humans are known to learn \citep{lillicrap2020backpropagation}. Generalizing layer-local learning to multi-layer objective functions has been the focus of much recent work: many multi-layer models can be seen as generalizations of perceptron learning \citep{bengio2014auto, hinton1995wake, rao1999predictive}, with other models such as those derived from similarity matching \citep{pehlevan2017similarity} receiving similar treatment \citep{obeid2019structured}. We will refer to this form of multi-layer signal propagation as `spatial' credit assignment, and will refer to relaying information across time as `temporal' credit assignment (Fig. \ref{fig_2}b; Section \ref{temporal}). As we will discuss in the next section, models that do not support temporal credit assignment will not be able to account for learning in inherently sequential tasks.

\subsection{Temporal credit assignment}\label{temporal}
Because so many learned biologically-relevant tasks involving temporal decision-making \citep{gold2007neural} or working memory \citep{compte2000synaptic, wong2006recurrent, ganguli2008memory} inherently leverage information from the past to inform future behavior, and because neural signatures associated with these tasks exhibit rich recurrent dynamics \citep{brody2003timing, shadlen2001neural, mante2013context, sohn2019bayesian}, many aspects of learning in the brain require a normative theory of synaptic plasticity that works in recurrent neural architectures and provides an account of temporal credit assignment.

Temporal credit assignment is an important point of failure of modern deep learning methods, in part due to the inherent instabilities involved in performing gradient descent on recurrent neural architectures \citep{bengio1994learning}. That models unconstrained in their correspondence to biology have difficulties handling temporal signals should be some indication of the difficulties posed by temporal credit assignment for normative theories of synaptic plasticity. However, recent improvements in neural architectures, including gated recurrent units \citep{chung2014empirical} and long short-term memory units \citep{hochreiter1997long}, as well as sequential reinforcement learning methods \citep{mnih2015human, arjona2019rudder, hung2019optimizing, raposo2021synthetic}, have combined to produce several high-profile advances on inherently temporal, naturalistic tasks like game-playing \citep{silver2017mastering} and natural language processing \citep{devlin2018bert, radford2018improving}. This may indicate that the time is ripe to begin incorporating new developments in deep learning into normative plasticity models.

As it currently stands, the majority of normative synaptic plasticity models focus only on spatial credit assignment, which presents distinct challenges when compared to temporal credit assignment \citep{marschall2020unified}. In fact, many theories that provide a potential solution to spatial credit assignment do so by requiring networks to relax to a `steady-state' on a timescale much faster than inputs \citep{hopfield1982neural, scellier2017equilibrium, bredenberg2020learning, xie2003equivalence, ackley1985learning}, which effectively prevents networks from having the rich, slow internal dynamics required for many temporal motor \citep{hennequin2012non} and working memory \citep{wong2006recurrent} tasks. Other methods appear to be agnostic to the temporal properties of their inputs, but have not yet been combined with existing plasticity rules that perform approximate temporal credit assignment within local microcircuits \citep{murray2019local, bellec2020solution}.

While most normative theories focus on spatial credit assignment, some new algorithms do provide potential solutions to temporal credit assignment, through either explicit approximation of real time recurrent learning \citep{marschall2020unified, bellec2020solution, murray2019local}, by leveraging principles from control theory \citep{gilra2017predicting, alemi2018learning, meulemans2022least}, or by leveraging principles of stochastic circuits that are fundamentally different from traditional explicit gradient-based calculation methods \citep{bredenberg2020learning, miconi2017biologically}. Many use what is called an `eligibility trace' \citep{gerstner2018eligibility} (Fig. \ref{fig_2}b)---a local synaptic record of coactivity---to identify associations between rewards and neural activity that may have occurred much further in the past. We suggest that these models capture something fundamental about learning across time, and that much work remains to combine these with spatial learning rules to construct normative models of full spatiotemporal learning.

\subsection{Combining learning and active performance} \label{online}
Similar to the importance of understanding temporal credit assignment in the brain, it is critical to understand how learning in the brain relates to continuous action and perception in an environment (Fig. \ref{fig_2}b). The relationship between learning and active performance in the brain can vary widely depending on the experimental context: learning-related changes can occur concomitantly with action \citep{bittner2015conjunctive, sheffield2017increased, grienberger2022entorhinal} (`online' learning), during brief periods of quiescence between trials \citep{pavlides1989influences, bonstrup2019rapid, liu2021experience}, or over periods of extended sleep \citep{gulati2017neural, eschenko2008sustained, girardeau2009selective} (`offline' learning). Therefore, whether a normative plasticity model uses offline or online learning should be determined by the experimental context, be it for instance rapid place cell reorganization in new environments, or long timescale memory consolidation.

However, many classical algorithms---especially those that support multi-layer spatial credit assignment \citep{ackley1985learning, xie2003equivalence, dayan1995helmholtz}---are constrained to modeling only offline learning, because they require distinct training phases, during at least one phase of which activity of neurons is driven for \textit{learning}, rather than performative purposes. It has not been clear whether such algorithms are fundamentally offline, or whether the space of phenomena that they can model can be expanded until recently.
Some existing two-phase normative algorithms, such as the Wake-Sleep algorithm (Appendix \ref{wake_sleep_tutorial}) \citep{hinton1995wake, dayan1995helmholtz}, have be adapted such that the second phase becomes indistinguishable from perception \citep{bredenberg2020learning, ernoult2020equilibrium}. Other recent models allow for simultaneous multiplexing of top-down learning signals and bottom-up inputs \citep{payeur2021burst}, which enables online learning. These results suggest that future work may fruitfully adapt existing offline algorithms to provide good models of explicitly online learning in the brain.

\subsection{Scaling in dimensionality and complexity} \label{scalable}
A point often underappreciated in computational neuroscience (and possibly overappreciated in machine learning) is that models of learning in the brain need to be able to scale to handle the full complexity of the problems a given model organism has to solve. As obvious as this sounds, it is a point that can be difficult to verify: how can we guarantee that adding more neurons and more complexity will not make a particular collection of plasticity rules more effective? As a case study, consider REINFORCE (\citep{williams1992simple}; Appendix \ref{reinforce_tutorial}), an algorithm which, for the most part, satisfies our other desiderata for normative plasticity for the limited selection of tasks in naturalistic environments which are explicitly rewarded. However, though REINFORCE demonstrably performs better than its progenitor weight perturbation \citep{jabri1992weight}, as the dimensionality of its stimuli, the number of neurons in the network, and the delay time between neural activity and reward increases, the performance of the algorithm decays rapidly, both analytically and in simulations \citep{werfel2003learning}. This is primarily caused by high variance of gradient estimates provided by the REINFORCE algorithm, and is only partially ameliorated by existing methods that reduce its variance \citep{bredenberg2021impression, ranganath2014black, mnih2014neural, miconi2017biologically}. Thus, adding additional complexity to the network architecture actually \textit{impairs} learning.

We do not mean to imply that all normative plasticity algorithms should be demonstrated to meet human-level performance, or even that they should match state-of-the-art machine learning methods. Machine learning methods profit in many ways from their biological implausibility: they use stochastic backpropagation, which is demonstrably biologically implausible (Appendix \ref{weight_transport}) but which benefits from very low variance gradient estimates \citep{werfel2003learning}; they share weights across topographically distant space in convolutional neural networks \citep{fukushima1982neocognitron}; they use rate-based units, which generally perform better than spiking units \citep{neftci2019surrogate}; and they are usually deterministic, which obviates the need for redundancy (increased neuron numbers) and increased computational demand. Beyond machine learning methods, the human brain itself has orders of magnitude more neural units and synapses than have ever been simulated on a computer, all of which are capable of processing totally in parallel. Therefore, direct comparison to the human---or any---brain is also not fair. We propose the far softer condition that as the complexity of input stimuli and tasks increase, within the range supported by current computational power, plasticity rules derived from normative theory should continue to perform well both in simulation and, preferably, analytically. Further, the performance of normative plasticity algorithms can fruitfully be compared to existing machine learning methods as long as the comparison is performed for realistic network architectures with identical conditions, as in \citep{bredenberg2021impression, payeur2021burst, marschall2020unified, bartunov2018assessing}.

Complexity is multifaceted, and involves features of both stimulus and task (Fig. \ref{fig_2}c). Even stimuli with very high dimensional structure can fail to capture critical features of naturalistic stimuli, as evidenced by the wide gap in difficulty involved in constructing convincing models that synthesize images with low-level naturalistic features (orientation, contrast, texture \citep{portilla2000parametric}) compared to models that capture high-level image features (object identity \citep{rezende2014stochastic, goodfellow2014generative}, semantic content \citep{ramesh2021zero}), which are only just beginning to emerge. Algorithms that scale well with the dimensionality of a stimulus can fail to capture high-level stimulus features: for example, PCA-based image models are unable to capture natural image statistics, and do not result in realistic neural receptive field properties \citep{olshausen1996emergence}. For these reasons, it is critical that normative plasticity algorithms be able to scale not just to high-dimensional `toy' datasets, but also to complex naturalistic data across sensory modalities. This is a major avenue for improvement: for instance, existing plasticity models have great difficulty scaling to naturalistic image datasets \citep{bartunov2018assessing}.

Similarly, naturalistic task structures are often much more complex than those used for training general machine learning algorithms, let alone models of normative plasticity (Fig. \ref{fig_2}c). In natural environments, rewards are often provided after long sequences of complex actions, supervised feedback is sparse, if present at all, and an organism's self preservation often requires navigating both uncertainty and complex multi-agent interactions. Modern reinforcement learning algorithms are only just beginning to make progress with some of these difficulties \citep{kaelbling1998planning, arjona2019rudder, raposo2021synthetic, hung2019optimizing, zhang2021multi}, but as yet there are no normative plasticity models that describe how any of the human capabilities used to solve these problems could be learned through cellular adaptation (for example, model-based planning \citep{doll2012ubiquity}); similarly, none of these capabilities have been shown to be an emergent consequence of a more basic plasticity process.

\subsection{Generating testable predictions} \label{testability}
\begin{figure}[t!]
	\includegraphics[width=\linewidth]{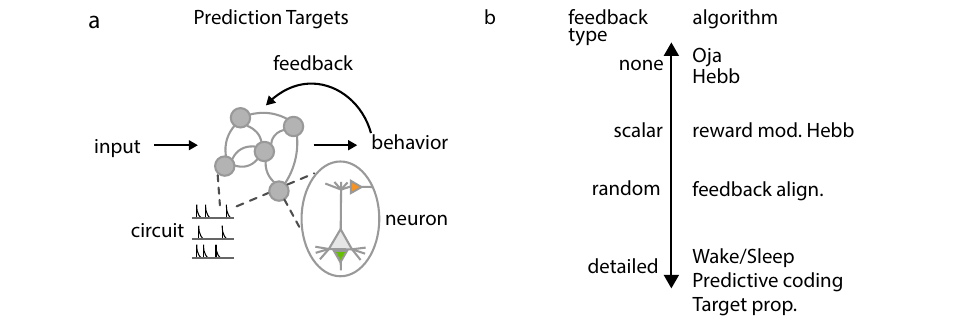}
    \centering
	\caption{ \small {\bf Testing normative theories.} {\bf a.} Normative plasticity theories can be assessed through four different experimental lenses centered on individual neurons, circuits of collectively recorded neurons, the training signals delivered to a circuit, and the organism's overall behavior over the course of learning. {\bf b.} Different normative plasticity theories postulate different levels of detail for the feedback signals received by individual neurons.}
	\label{fig_3}
\end{figure}
Despite the abundance of existing normative theories, very few have been confirmed experimentally, and of those that have received partial confirmation, they are restricted to very specific experimental preparations, for example: fear conditioning in \textit{Aplysia} \citep{rayport1986synaptic}, and reward-based learning in songbird motor systems \citep{fiete2007model} and in mouse auditory cortex \citep{froemke2013long, guo2019cholinergic}. This relative paucity of validation will not be overcome without a very clear articulation of which features of a normative theory constitute testable predictions, and in what way those predictions disambiguate one theory from its alternatives.

Many existing features of normative theories would be fatal to those theories if proven not to hold in biology. Some examples include: weight symmetry, reward modulation of plasticity, differential roles (and plasticity rules) for apical and basal synapses, and the existence of eligibility traces for temporal credit assignment. However, these individual features, if \textit{proven} to hold, would eliminate alternative theories to highly variable degrees. Most, if not all models could accommodate weight symmetry, several distinct models predict reward modulation of plasticity either through precise credit assignment or global neurotransmitter delivery \citep{murray2019local, williams1992simple, bellec2020solution, roth2018kernel}, and several distinct supervised and unsupervised models predict different types of signaling and plasticity at apical and basal synapses on pyramidal neurons \citep{urbanczik2014learning, payeur2021burst, bredenberg2021impression, kording2001supervised, schiess2016somato, sacramento2017dendritic, guerguiev2017towards, richards2019dendritic}, while nearly all models capable of temporal credit assignment assume some form of synaptic eligibility trace \citep{bellec2020solution, marschall2020unified, murray2019local, miconi2017biologically, roth2018kernel}. It is intuitively clear that for any given normative theory of synaptic plasticity, there exist an infinite number of infinitesimal perturbations to that theory that would be impossible to disambiguate experimentally. Further, there are many features of normative theories that would be fatal if proven not to hold, but are completely unclear how to test experimentally.

The most useful predictions are those that are fatal to the theory if proven false, are clearly testable, and disambiguate the theory from the greatest number of alternative theories. It may be that a collection of predictions is required to completely isolate one individual normative theory from closely related models, which suggests that articulating where particular models lie within a taxonomy of predictions is the most useful way to narrow down the field of possible models. Testable predictions can be defined in terms of several different experimental lenses, of which we isolate four: experiments examining individual neurons or synapses, populations of neurons, the feedback mechanisms that shape learning in neural circuits, or learning at a behavioral level (Fig. \ref{fig_3}a). Accurately distinguishing one mechanism from another will likely require a synthesis of experiments spanning all four lenses.

{\bf Individual neurons.}
Experiments that focus on individual neurons, including paired-pulse stimulation \citep{markram1997regulation}, mechanistic characterizations of plasticity \citep{graupner2010mechanisms}, pharmacological explorations of neuromodulators that induce or modify plasticity \citep{bear1986modulation, reynolds2002dopamine, froemke2007synaptic, gu1995involvement}, and characterization of local dendritic or microcircuit properties mediating plasticity \citep{froemke2005spike, letzkus2006learning, sjostrom2006cooperative} form the bulk of the classical literature underlying phenomenological and mechanistic modeling. These studies characterize what information is locally available at synapses and what can be done with that information, as well as which properties of cells can be altered in an experience-dependent fashion.

Existing normative theories differ in the nature of their predictions for plasticity at individual neurons. Reward-modulated Hebbian theories \textit{require} feedback information be delivered by a neuromodulator like dopamine, serotonin, or acetylcholine \citep{fremaux2016neuromodulated} and that this feedback modulates plasticity at the local synapse by changing the magnitude or sign of plasticity depending on the strength of feedback. In contrast, some unsupervised normative theories require no feedback modulation of plasticity \citep{pehlevan2015hebbian, pehlevan2017similarity}, and others argue that detailed feedback information arrives at the apical dendritic arbors of pyramidal neurons to modulate plasticity, which is also partially supported in the hippocampus \citep{bittner2015conjunctive, bittner2017behavioral} and cortex \citep{larkum1999new, letzkus2006learning, froemke2005spike, sjostrom2006cooperative}.

Independent of the exact feedback mechanism, models differ in how temporal associations are formed. Algorithms related to REINFORCE assume that local synaptic eligibility traces integrate over time fluctuations in coactivity of the post- and pre-synaptic neuron local to a synapse. These postulated eligibility traces are stochastic, summing Gaussian fluctuations in activity \citep{miconi2017biologically} that consequently produce temporal profiles similar to Brownian motion. In contrast, methods based on approximations to real-time recurrent learning propose eligibility traces that are deterministic records of coactivity whose time constants are directly connected to the dynamics of the neuron itself \citep{bellec2020solution}, while other hybrid approaches predict eligibility traces which are deterministic but are related more to predicted task timescale than the dynamics of the cell \citep{roth2018kernel}. Though there do exist known cellular processes that naturally track coactivity, like NMDA receptors \citep{bi1998synaptic}, and that store traces of this coactivity longitudinally, like CaMKII \citep{graupner2010mechanisms}, much work remains to be done to analyze how the properties of these known biophysical quantities relate to the predictions of various normative theories, and whether there are other biological alternatives. Other algorithms have different predictions at a microcircuit, rather than at an individual neuron level. Impression learning, for instance, suggests that a population of inhibitory interneurons could gate the influence of apical and basal dendritic inputs to the activity of pyramidal neurons \citep{bredenberg2021impression}, and some forms of predictive coding propose that top-down error signals are partially computed by local inhibitory interneurons. Therefore, to completely distinguish different theories, it may be necessary to analyze the connectivity and plasticity between small groups of different cell types.

In sum, experiments at the level of individual neurons or local microcircuits potentially have a great deal of power to identify whether a particular neural circuit is implementing any of a collection of hypothesized normative models of plasticity. It is an advantage that these methods can identify the adaptive capabilities of individual neurons and synapses, but these methods are also limited in their ability to simultaneously observe the adaptation of many neurons in a circuit. Normativity is inherently concerned with the value of plasticity for perception and behavior, and as we will see in subsequent sections, experiments targeting larger populations of neurons will be necessary to distinguish certain features of these theories.

{\bf Neural circuits.}
Determining how circuits encode environmental information and affect motor actions by an animal cannot be assessed by looking at single neurons, and by extension, analyzing how these properties change over time requires methods that record large groups of neurons, such as 2 photon calcium imaging, multielectrode recordings, fMRI, EEG, and MEG, as well as methods that manipulate large populations, like optogenetic \citep{rajasethupathy2016targeting} stimulation.
The benefits of these recording techniques for testing normative plasticity models, though less practiced compared to individual neuron studies, are manyfold. One of the challenges for characterizing a circuit with a normative plasticity model is selecting an appropriate objective function. Determining which objective fits best can partly be determined by philosophical considerations (Section \ref{objective}), but empirical validation is a far more rigorous test. For instance, one can establish that explicit reward modifies a neural representation to improve coding of task-relevant variables \citep{froemke2013long}. Another line of approaches trains neural networks on a battery of objectives, and determines which objective produces the closest correspondence between model neurons and neurons recorded brain in a variety of areas in the ventral \citep{yamins2014performance, yamins2016using} and dorsal \citep{mineault2021your} visual streams, as well as recently in auditory cortex \citep{kell2018task} and medial entorhinal cortex \citep{nayebi2021explaining}. Oftentimes, changes in artificial neural network activity throughout time are sufficient to determine the objective optimized by the network as well as its learning algorithm \citep{nayebi2020identifying}, an approach which could also potentially be applied to recorded neural activity over learning.

Beyond narrowing down the objective function, recording from populations can establish features of neural learning that normative models must account for. For instance, in biofeedback training settings, animals can selectively control the firing rates of individual neurons to satisfy arbitrary experimental conditions for reward \citep{fetz2007volitional}, suggesting the existence of highly flexible credit assignment systems in the brain, which are not constrained by evolutionary predetermination of the function of neural circuits\footnote{This is a challenge for normative plasticity models that predefine the outputs of the circuit and approximately backpropagate errors from these outputs.}.
Further, circuit recordings could in principle test predictions about how neural circuits should function in situations that do not specifically involve learning. For instance, the Wake-Sleep algorithm \citep{dayan1995helmholtz} (Appendix \ref{wake_sleep_tutorial}) proposes that neural circuits should spend extended periods of time (e.g. during dreaming) generating similar activity patterns to those evoked by natural stimulus sequences, whereas impression learning proposes that similar hallucinatory states could be induced by experimentally increasing the influence of apical dendrites on pyramidal neuron activity \citep{bredenberg2021impression}. An alternative learning algorithm based on generative adversarial networks proposes that during sleep networks rehearse corrupted versions of recent waking experiences \citep{deperrois2021memory}. There is plenty of room for experiments to more clearly map predictions and components of these models onto well documented neural phenomena, such as sleep or potentially replay phenomena \citep{girardeau2009selective, eschenko2008sustained}.
Because circuit recording and manipulation methods often sacrifice temporal resolution \citep{hong2019novel}, and have difficulty inferring biophysical properties of individual synapses and cells, these methods are best used in concert with single neuron studies to jointly tease apart the multi-level predictions of various normative models.

{\bf Feedback mechanisms.}
One of the best ways to distinguish normative plasticity algorithms is on the basis of the nature of their feedback mechanisms (Fig. \ref{fig_3}b). Though some unsupervised algorithms, like Oja's rule propose that no feedback is necessary to perform meaningful learning, no current normative theories propose any form of supervised or reinforcement learning that does not require \textit{some} form of top-down feedback. However, across these models, the level of precision of feedback varies considerably. The simplest feedback is scalar, conveying reward \citep{williams1992simple}, state fluctuation \citep{payeur2021burst}, or context (e.g. saccade \citep{illing2021local} or attention \citep{roelfsema2005attention, pozzi2020attention}) information. Beyond this, the space of proposed mechanisms expands considerably: backpropagation approximations like feedback alignment \citep{lillicrap2016random} and random-feedback online learning (RFLO) \citep{murray2019local} propose random feedback between layers of neurons can provide a sufficient learning signal, whereas algorithms based on control theory propose that low-rank or partially random projections carrying supervised error signals are sufficient \citep{gilra2017predicting, alemi2018learning}. Other algorithms propose even more detailed feedback, with individual neurons receiving precise, carefully adapted projections carrying learning-related information. These algorithms propose that top-down projections to apical dendrites \citep{urbanczik2014learning} or local interneurons neurons \citep{bastos2012canonical} perform spatial credit assignment, but the nature of this signal can differ considerably across different algorithms. It could be a supervised target, carrying information about what the neuron state `should' be to achieve a goal \citep{guerguiev2017towards, payeur2021burst}, or it could be a prediction of the future state of the neuron \citep{bredenberg2020learning}.
        %

Each of these different possibilities is theoretically testable, if the focus is shifted to the postulated feedback mechanism, instead of the circuit undergoing learning. However, so far the different mechanisms have received only partial support. For example, acetylcholine projections to auditory cortex that modulate perceptual learning \citep{froemke2013long} display a diversity of responses related to both reward and attention \citep{hangya2015central}, which adapt over the course of learning in concert with auditory cortex \citep{guo2019cholinergic}. This suggests that while traditional models of reward-modulated Hebbian plasticity may be correct to a first approximation, a more detailed study of the adaptive capabilities of neuromodulatory centers may be necessary to update the theories.

While a growing number of studies indicate that projections to apical synapses of pyramidal neurons \textit{do} play a role in inducing plasticity, and that these projections themselves are also plastic (i.e. nonrandom) \citep{bittner2015conjunctive, bittner2017behavioral}, very little is known about the \textit{nature} of the signal---a critical component for distinguishing several different theories. In the visual system, presentation of unfamiliar images without any form or reward or supervision can modify both apical and basal dendrites throughout time \citep{gillon2021learning}, and in the hippocampus, apical input to CA1 pyramidal neurons while animals acclimatize to new spatial environments is sufficient to induce synaptic plasticity \citep{bittner2015conjunctive, bittner2017behavioral}. These two examples support a form of \textit{unsupervised} learning, but evidence for supervised or reinforcement learning signals propagated through apical dendritic synapses is currently lacking. Beyond the cerebellar system, where climbing fiber pathways may carry explicit motor error signals used for plasticity \citep{gao2012distributed, bouvier2018cerebellar}, evidence for detailed supervised feedback is limited.
In sum, beyond single neurons, or even populations recorded by traditional techniques, targeted focus on the learning feedback signals received by a population shows promise to rule out algorithms on the basis of their feedback and objective function.

{\bf Behavior.}
In much the same way that psychophysical studies of human or animal responses define constraints on what the brain's perceptual systems are capable of, behavioral studies of learning can do quite a lot to describe the range of phenomena that a model of learning must be able to capture, from operant conditioning \citep{niv2009reinforcement}, to model-based learning \citep{doll2012ubiquity}, rapid language learning \citep{heibeck1987word}, unsupervised sensory development \citep{wiesel1963single}, or consolidation effects \citep{stickgold2005sleep}. Behavioral studies can also outline key limitations in learning, which are perhaps reflective of the brain's learning algorithms, including the brain's failure to perform certain types of adaptation after critical periods of plasticity \citep{wiesel1963single}, and the brain's unexpected inability to learn multi-context motor movements without explicit motor differences across contexts \citep{sheahan2016motor}.

These existing experimental results stand as (often unmet) targets for normative theories of plasticity, but in addition, normative theories themselves suggest further studies that may test their predictions. In particular, manipulation of learning mechanisms may have predictable effects on animals' behavior, as seen when acetylcholine receptor blockage in mouse auditory cortex prevented reward-based learning in animals \citep{guo2019cholinergic}, and nucleus basalis stimulation during tone perception longitudinally improved animals' discrimination of that tone \citep{froemke2013long}. Other algorithms have as-yet untested predictions for behavior: for instance, experimentally increasing the influence of top-down projections should bias behavior towards commonly-occurring sensory stimuli according to both predictive coding \citep{rao1999predictive, friston2010free} and impression learning \citep{bredenberg2021impression}. For other detailed feedback algorithms (Fig. \ref{fig_3}b), manipulating top-down projections may disrupt learning, but would have a much more unstructured deleterious effect on perceptual behavior.

As shown, each experimental lens has its own advantages and disadvantages. Single-neuron studies are excellent for identifying the locally-available variables that affect plasticity, circuit-level studies can help narrow down the objectives that shape neural responses and identify traces of offline learning, studies of feedback mechanisms can distinguish between different algorithms that postulate different degrees of precision in their feedback and in complexity of the teaching signal, and studies of behavior can place boundaries on what can be learned, as well as serve as a readout for manipulations of the mechanisms underlying learning. Each focus alone is insufficient to distinguish between all existing normative models, but in concert they show promise for identifying the neural substrates of adaptation.
\setcitestyle{numbers}
\defcitealias{werbos1974beyond}{Wer74}
\defcitealias{williams1992simple}{Wil92}
\defcitealias{lee2016training}{Lee16}
\defcitealias{werbos1990backpropagation}{Wer90}
\defcitealias{miconi2017biologically}{Mic17}
\defcitealias{werfel2003learning}{Wer03}
\defcitealias{oja1982simplified}{Oja82}
\defcitealias{rao1999predictive}{Rao99}
\defcitealias{whittington2017approximation}{Whi17}
\defcitealias{friston2009cortical}{Fri09}
\defcitealias{dayan1995helmholtz}{Day95}
\defcitealias{dayan1996varieties}{Day96}
\defcitealias{bredenberg2021impression}{Bre21}
\defcitealias{lillicrap2016random}{Lil16}
\defcitealias{akrout2019using}{Akr19}
\defcitealias{bellec2020solution}{Bel20}
\defcitealias{murray2019local}{Mur19}
\defcitealias{scellier2017equilibrium}{Sce17}
\defcitealias{ernoult2020equilibrium}{Ern20}
\defcitealias{laborieux2021scaling}{Lab21}
\defcitealias{bengio2014auto}{Ben14}
\defcitealias{manchev2020target}{Man20}
\defcitealias{lee2015difference}{Lee15}
\begin{table}[t!]
\centering
\small
\begin{tabular}{|c|| c | c |c|c|c|c|c|}
    \hline
     Algorithm & Dec. Loss & Local  &  Arch. & Time & Online & Scalable  \\
     \hline
     \hline
     Backpropagation  &  U/S/R  & \xmark & \cmark  & \cmark  &  \xmark & \cmark \\
     \citepalias{werbos1974beyond} & \citepalias{williams1992simple} & & \citepalias{lee2016training} & \citepalias{werbos1990backpropagation} & & \\
     \hline
     REINFORCE &  U/S/R & \cmark & \cmark & \cmark &  \cmark & \xmark \\
     \citepalias{williams1992simple} & & & & \citepalias{miconi2017biologically} & & \citepalias{werfel2003learning}\\
     \hline
     Oja \citepalias{oja1982simplified} & U & \cmark &  \xmark & \xmark & \cmark & \cmark\\
     \hline
     Predictive Coding  & U/S  & \cmark &  \xmark & \cmark  &\cmark & \cmark \\
     \citepalias{rao1999predictive} & \citepalias{whittington2017approximation} & & &  \citepalias{friston2009cortical} & &  \\
     \hline
     Wake-Sleep & U & \cmark &  \cmark & \cmark  & \cmark & \cmark\\
     \citepalias{dayan1995helmholtz} & & & \citepalias{dayan1996varieties} & \citepalias{dayan1996varieties} & \citepalias{bredenberg2021impression} & \\
     \hline
     Approx. Backprop. & U/S* & \cmark  &  \cmark  & \cmark  & \cmark  & \cmark \\
     \citepalias{lillicrap2016random} & & & \citepalias{bellec2020solution} & \citepalias{murray2019local} & \citepalias{murray2019local} & \\
     \citepalias{akrout2019using} & & & & \citepalias{bellec2020solution} & \citepalias{bellec2020solution} &\\
     \hline
     Equilibrium Prop. & U/S & \cmark &  \xmark & \xmark & \cmark & \cmark  \\
     \citepalias{scellier2017equilibrium} & & & & & \citepalias{ernoult2020equilibrium} & \citepalias{laborieux2021scaling} \\
     \hline
     Target Prop.  & U/S & \cmark & \cmark & \cmark  &  \xmark & \cmark  \\
     \citepalias{bengio2014auto} & & & & \citepalias{manchev2020target} & & \citepalias{lee2015difference} \\
     \hline
    
\end{tabular}
\caption{ \small {\bf Summarizing progress on the desiderata.} A \cmark~ indicates that an algorithm has been demonstrated to satisfy a particular desideratum in at least one study, whereas an \xmark~ indicates that it has not been demonstrated. If the demonstrating study is an improvement on the seminal work or is a new model, we provide a citation;  reference numbering used for brevity: Asterisks (*) indicate that results have only been shown by simulation, and lack mathematical support. U, S, and R indicate whether a given algorithm supports unsupervised, supervised, or reinforcement learning, respectively.
} \label{table_1}
\end{table}
\setcitestyle{authoryear}
\section{Conclusions}
Normative plasticity models are compelling because of their potential to connect our brains' capacity for adaptation to their constituent synaptic modifications. 
Generating good theories is a critical part of the scientific process, but finding ways to close the loop by testing key predictions of new normative models has proved extraordinarily difficult: in this perspective we have illustrated the sources of this difficulty. 

The core of a normative plasticity model is its plasticity rule, which dictates how a model synapse modifies its strength. To be a normative model---to explain why the plasticity mechanism is important for the organism---there must be a concrete demonstration that this plasticity rule supports adaptation critical for system-wide goals like processing sensory signals or obtaining rewards (Section \ref{objective}). However, this system-wide goal must be achieved using only \textit{local} information (Section \ref{locality}). These two needs of a normative plasticity model are the fundamental source of tension: it is very difficult to demonstrate that a proposed plasticity rule is both local \textit{and} optimizes a system-wide objective (Appendix \ref{weight_transport}). Insufficient or partial resolution of this fundamental tension produces normative models that struggle to map accurately onto neural hardware (Section \ref{architecture}) or handle complex temporal stimuli and tasks online (Sections \ref{temporal}-\ref{scalable}). To provide a case study of how our desiderata come to be satisfied (or not) in practice, we have included tutorials for both REINFORCE and the Wake-Sleep algorithm in Appendices \ref{reinforce_tutorial} and \ref{wake_sleep_tutorial}. These tutorials are by no means a complete introduction to the field, but will hopefully serve as a solid foothold for analyzing modern normative plasticity models.

Even satisfying the aforementioned desiderata, much work remains to delineate which tests would most clearly distinguish a normative model from its alternatives in a biological system.
In this review, we have organized emerging theories according to how they satisfy and improve upon our desiderata (Table \ref{table_1}), as well as by how they can be tested (Section \ref{testability}), with the view that this organization will provide avenues for both experimental and theoretical neuroscientists to bring normative plasticity models closer to biology. Even if existing algorithms prove not to be implemented exactly in the brain, they undoubtedly provide key insights into how local synaptic modifications can produce valuable improvements in both behavior and perception for an organism. It seems sensible to use these algorithms as a springboard to produce more biologically realistic and powerful theories.

Beyond improving normative theories with respect to our desiderata, there are several incredible opportunities for actually testing their implementation in biology (Section \ref{testability}). Most current theoretical studies of reward-modulated Hebbian plasticity focus on dopamine-modulated motor learning in monkeys and songbirds \citep{fiete2007model, legenstein2010reward}, but there are \textit{many} neuromodulatory systems that have been linked to learning in experiments, including serotonin-modulated fear conditioning in the amygdala \citep{lesch2012serotonin}, as well as acetylcholine-modulated reward learning and oxytocin-modulated social learning in mouse auditory cortex \citep{guo2019cholinergic, froemke2013long}. Further, several experimental preparations examine the relationship between pyramidal neurons' apical and basal dendritic activity and plasticity, in both the hippocampus \citep{bittner2015conjunctive, bittner2017behavioral} and visual cortex \citep{gillon2021learning, froemke2005spike, letzkus2006learning, sjostrom2006cooperative}. These could test at the level of individual neurons, circuits, behavior, and the feedback mechanisms that support plasticity, which of the many alternative normative theories underlie animals' learning. 

As the diversity of aforementioned experimental preparations suggests, there are increasingly strong arguments for several fundamentally different plasticity algorithms instantiated in different areas of the brain and across different organisms, subserving different functions. It is quite likely that many plasticity mechanisms work in concert to produce learning as it manifests in our perception and behavior. It is our belief that well-articulated normative theories can serve as the building blocks of a conceptual framework that tames this diversity and allows us to understand the brain's tremendous capacity for adaptation.


\section{Acknowledgements and Funding}
We would like to thank Blake Richards, Eero Simoncelli, Owen Marschall, Benjamin Lyo, Elliott Capek, Olivier Codol, and Yuhe Fan for their helpful feedback on this manuscript. CS is supported by NIMH Award 1R01MH125571-01, NIH Award R01NS127122, by the National Science Foundation under NSF Award No. 1922658 and a Google faculty award.

\bibliographystyle{apalike}
\bibliography{desiderata}

\newpage
\begin{appendix}
\setcounter{figure}{0}
\makeatletter 
\renewcommand{\thefigure}{S\@arabic\c@figure}
\makeatother

\section{The unidentifiability of an objective} \label{identifiability}
In this section we illustrate why the choice of objective function for a normative plasticity model is never uniquely determined by data. We will consider two situations: the system has already settled to its optimal setting of its weights, $\weight^*$, and in the second we are able to observe the system's plasticity update $\Delta \weight$.

\subsection{Unidentifiability based on an optimum}
Suppose that some setting of synaptic weights $\weight^*$ minimizes an objective function $\mathcal{L}$, i.e. $\mathcal{L}(\weight^*) \leq \mathcal{L}(\weight) ~ \forall \weight$. We might be tempted to argue that because $\weight^*$ minimizes $\mathcal{L}$, $\mathcal L$ must be \textit{the} objective that the system is minimizing. However, there are an infinite variety of alternative objectives that share the same minimum. To see this, take a new objective $\tilde{\mathcal{L}} = \sigma \left (\mathcal{L}(\weight) \right )$ for any differentiable, monotonically increasing function $\sigma(\cdot)$. Then we have:

\begin{align}
    &\mathcal{L} (\weight^*) \leq \mathcal{L} (\weight) ~ \forall \weight \\
    \Rightarrow &\sigma \left (\mathcal{L} (\weight^*) \right )\leq \sigma \left (\mathcal{L} (\weight) \right ) ~ \forall \weight\\
    \Rightarrow &\tilde{\mathcal{L}} (\weight^*) \leq \tilde{\mathcal{L}} (\weight) ~ \forall \weight,
\end{align}
where the second equality follows from the order preservation property of $\sigma(\cdot)$. This means that $\weight^*$ also minimizes $\tilde{\mathcal{L}}$, i.e. we will be unable to arbitrate between whether the system is `attempting' to minimize $\tilde{\mathcal{L}}$ or $\mathcal{L}$ on the basis of the optimized network state given by $\weight^*$.

\subsection{Unidentifiability based on an update rule}
Suppose instead that we were able to observe the adaptive plasticity mechanism of a system, and were able to verify that it really does decrease an objective function $\mathcal{L}$, i.e. by Eq. \ref{inner_product},

\begin{equation}
    \derivative{\mathcal{L}}{\weight}(\weight)^T \Delta \weight \leq 0 ~ \forall \weight.
\end{equation}

We might now be tempted to argue that, by observing the plasticity rule itself, $\Delta \weight$, we will be more able to assert that the system, by virtue of consistently decreasing $\mathcal{L}$, is `attempting' to minimize $\mathcal{L}$. However, the \textit{exact same} family of alternative objectives will also be minimized ($\tilde{\mathcal{L}} = \sigma \left (\mathcal{L}(\weight) \right )$ for any differentiable, monotonically increasing function $\sigma(\cdot)$). To see this, we observe:

\begin{align}
    &\derivative{\mathcal{L}}{\weight}(\weight)^T \Delta \weight \leq 0 ~ \forall \weight \\
    \Rightarrow &\derivative{\sigma(\mathcal{L}(\weight))}{\mathcal{L}(\weight)}\derivative{\mathcal{L}}{\weight}(\weight)^T \Delta \weight \leq 0 ~ \forall \weight \\
    \Rightarrow & \derivative{\tilde{\mathcal{L}}}{\weight}(\weight)^T \Delta \weight \leq 0 ~ \forall \weight,
\end{align}
where the first implication follows from the fact that $\sigma(\cdot)$ is differentiable and increasing (it has strictly positive derivative), and the second implication follows from the chain rule. This implies that plasticity rules ($\Delta \weight$) and trained neural circuits ($\weight^*$) can at most partially constrain the space of viable objective functions the system could be minimizing.

\section{Why can't the brain do explicit gradient descent?} \label{weight_transport}
\begin{figure}[t!]
	\includegraphics[width=\linewidth]{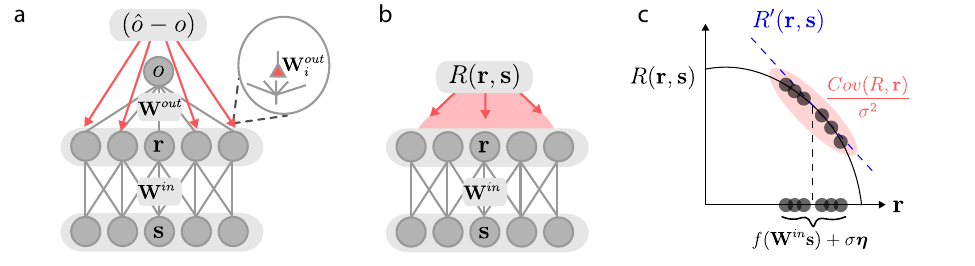}
    \centering
	\caption{\small {\bf Weight transport and REINFORCE.} {\bf a.} Traditional gradient descent propagates a credit assignment signal $(\target - \out)\decoder_i$ to each neuron $\rate_i$. How this pathway could have access to $\decoder_i$ is unclear: this is the `weight transport' problem. {\bf b}. REINFORCE resolves the weight transport problem by projecting a scalar reward signal $\reward$ to all synapses. {\bf c}. By correlating this reward with fluctuations in neural activity, neurons can approximate the true gradient.}
	\label{fig_supp_1}
\end{figure}
We have provided one surefire way to decrease an objective function by modifying the parameters of a neural network---`simply' take small steps in the direction of the gradient of the loss (Section \ref{objective}). To appreciate the challenges faced by theories of normative plasticity, it's important to understand why a biological system \textit{could not} do this: in this section we will provide a simplified argument as to why gradient descent within multilayer neural networks produces \textit{nonlocal} parameter updates, thus failing our most critical desideratum for a normative plasticity theory (Section \ref{locality}). More detailed arguments for multilayer neural networks can be found here \citep{lillicrap2020backpropagation}, and descriptions of why gradient descent becomes even more implausible for recurrent neural networks trained with either backpropagation through time \citep{werbos1990backpropagation} or real-time recurrent learning \citep{williams1989learning} can be found here \citep{marschall2020unified}.

The `weight transport problem' is the most basic reason that gradient descent is implausible for neural networks. Suppose that we have a stimulus-dependent network response, $\rate(\win) = f(\win \stim)$, where $\rate$ is an $N \times 1$ vector, and $\win$ is an $N \times N^s$ weight matrix mapping stimuli $\stim$ into responses after a pointwise nonlinearity $f(\cdot)$. This network response is decoded into a network output, $\out(\win, \stim) = \decoder \rate(\win)$, where $\decoder$ is a $1 \times N$ vector mapping network responses into a scalar output. Now suppose for simplicity that our loss for a single stimulus example is given by:

\begin{equation}
    \mathcal{L} = \frac{1}{2}\left ( \target - \out (\win, \stim) \right ) ^2.
\end{equation}

This objective is trying to bring the stimulus-dependent network response $\out (\win, \stim)$ close to the target output $\target$, and is zero if and only if $\out = \target$. A reasonable hypothesis would be that the gradient of this objective function with respect to a synaptic weight, $\win_{ij}$, will produce a parameter update that is local: we will see that this is not true. Taking the gradient, we have:

\begin{align}
    \derivative{}{\win_{ij}} \mathcal{L} &= \frac{1}{2}\derivative{}{\win_{ij}}\left ( \target - \out (\win, \stim) \right ) ^2 \\
    &= \left ( \target - \out\right ) \derivative{}{\win_{ij}} \out(\win, \stim)\\
    &= \left ( \target - \out\right ) \decoder_{i}  \derivative{}{\win_{ij}} f_i(\win \stim) \\
    &= \left ( \target - \out\right ) \decoder_{i} f'_i (\win \stim) \stim_j.
\end{align}

Breaking down this final update, we can see three terms: an error, $\left ( \target - \out \right )$, the neuron's \textit{output weight} $\decoder_{i}$, and an approximately Hebbian term $f'_i(\win \stim) \stim_j$, which requires only a combination of pre- and post-synaptic activity. One might be tempted to organize the plasticity rule into a error feedback signal received by the neuron, scaled by a neuron-specific synaptic weight $\decoder_{i}$, and then combined with Hebbian coactivity to produce a synaptic update (Fig. \ref{fig_supp_1}a).
This would have the form of a three-factor plasticity rule \citep{fremaux2016neuromodulated}, combining weighted feedback with pre- and post-synaptic activity. However, the weight transport problem is as follows: $\decoder_{i}$ provides the strength of a synapse in the \textit{feedforward} pathway---how could it possibly come to be that a feedback learning pathway would have access to the \textit{same} synaptic weight? The answer is that there is no evidence for such a system of weight sharing across feedforward and feedback pathways in the brain, though there are many hypotheses about how such a system could, in theory, be approximated by a normative plasticity algorithm. This problem becomes more pronounced in multilayer networks, where the error signal must be propagated through many interconnected connectivity layers.

It is also worth noting two key differentiability assumptions inherent to this approach. For one, we assume not only that the loss function $\mathcal{L}$ is differentiable, but that some `error calculating' part of the brain does differentiate it. This requires knowledge of what the desired network output should be $\target$, which for many real-world tasks is not possible. Second, we assume that the network activation function $f(\cdot)$ is differentiable. Since neurons typically emit binary spikes, this differentiability assumption is not necessarily valid, though several modern methods have circumvented this problem by using either stochastic neuron models \citep{williams1992simple, dayan1996varieties} or by using clever optimization tricks \citep{bellec2020solution}. In subsequent sections, we will outline two canonical algorithms that employ clever tricks to circumvent the weight transport problem.

\section{REINFORCE} \label{reinforce_tutorial}
In this section, we will provide a mathematical tutorial on the REINFORCE learning algorithm \citep{williams1992simple}, which is a mechanism for updating the parameters in a stochastic neural network for reinforcement learning objective functions. Its chief advantages are twofold: first, it only requires you to be able to evaluate an objective function (i.e. the reward received on any given trial), not the gradient of the objective function with respect to the parameters (Fig. \ref{fig_supp_1}b). This is very useful in situations in which the relationship between rewards and network outputs is not clear to an agent, as would be the case in many reinforcement learning scenarios. Second, under a broad range of biologically reasonable assumptions about a neural network architecture, the parameter updates produced by this algorithm are `local,' meaning the information required for a parameter update would reasonably be available to a synapse in the brain. This algorithm produces updates that are within the class of `reward-modulated Hebbian plasticity rules.'
The chief disadvantage of this algorithm is its comparative data-inefficiency relative to backpropagation. In practice, far more data samples (or equivalently, much lower learning rates) will be required to produce the same improvements in performance compared to backpropagation \citep{werfel2003learning}.

The REINFORCE algorithm and minor variations appear in different fields with different names. It is useful to keep track of these alternative names, because they all use roughly the same derivation, with some improvements or field-specific modifications.
In machine learning, the algorithm is often referred to as \textit{node perturbation} \citep{richards2019deep, lillicrap2020backpropagation, werfel2003learning}, because it involves correlating fluctuations in neuron (node) activity with reward signals. In computational neuroscience, it is sometimes called \textit{3-factor} or \textit{reward-modulated Hebbian} plasticity \citep{fremaux2016neuromodulated}, though REINFORCE is only one of several algorithms referred to by these blanket terms. In reinforcement learning, REINFORCE is often treated as a member of the more general class of \textit{policy gradient} \citep{sutton2018reinforcement} methods, which can be used to train any parameterized stochastic agent through reinforcement. Policy gradient methods need not commit to a neural network architecture, and are consequently not always local. Lastly, very similar methods are used for fitting variational Bayesian models, and are in these contexts referred to as either \textit{black box variational inference} \citep{ranganath2014black} or \textit{neural variational inference} \citep{mnih2014neural}.

In what follows, we will provide a brief derivation of the REINFORCE learning algorithm for a 1-layer feedforward neural network. We will then discuss the many extensions of the algorithm as well as its strengths and limitations as a normative plasticity model.

\subsection{Network model}

Most neural networks used in machine learning are deterministic. However, neurons in biological systems fluctuate across trials and stimulus presentations, so modeling them as stochastic is often more appropriate. It will turn out that these fluctuations can be used to produce parameter updates in a way that a deterministic system could not.

First, we will assume that there are stimuli drawn from some stimulus distribution, $p(\stim)$, and we will define the neural network response to a given stimulus drawn from this distribution as:

\begin{equation}
\rate = f(\win \stim) + \sigma \noise,
\end{equation}
where the $\noise$ is the source of random fluctuations which, for simplicity, is drawn from a standard normal distribution $(\mathcal{N}(0,1))$. In this equation, $\stim$ is an $N_s\times1$ vector, $\win$ is an $N_r \times N_s$ matrix, $f(\cdot)$ is the $\tanh$ nonlinearity, and $\noise$ is an $N_r \times 1$ vector.

This equation defines a conditional probability distribution, $p(\rate|\stim; \win) \sim \mathcal{N}(f(\win \stim), \sigma^2)$. There is an interesting point here: neuron activities are now samples from this conditional probability distribution, and so we can study how neurons behave on average by taking expectations over the probability distribution.

For simplicity and clarity we will restrict ourselves to this neural architecture for our derivation, but the basic principles apply more generally to a variety of noise sources and neural architectures (see Section \ref{assessing_reinforce}).

\subsection{Defining the objective}
We will assume that our goal is to maximize some instantaneous reward $\reward$ on average across many different samples of $\reward$ and $\stim$. This allows us to write our objective function $\obj$ as:

\begin{equation}
\obj = \int \reward p(\rate|\stim; \win) p(\stim) d\rate d\stim.
\end{equation}

In practice, this integral might be analytically impossible to integrate, but we can always approximate it (because it is an expectation) using samples from $p(\rate|\stim; \win)$ and $p(\stim)$ as an empirical average over $K$ samples $\rate_k$ and $\stim_k$:

\begin{equation}
\obj \approx \frac{1}{K} \sum_{k=0}^K R(\rate^{(k)}, \stim^{(k)}).
\end{equation}

Procedurally, this would amount to sampling $\stim$ and $\rate$ each $K$ times, calculating the reward for each trial, and taking an average.

\subsection{Taking the gradient}

Now that we have our objective function, we can evaluate its derivative with respect to a particular synapse $\win_{ij}$ in the network:

\begin{align}\label{reinforce_update_1}
\derivative{\obj}{\win_{ij}} &= \derivative{}{\weight} \int \reward p(\rate|\stim; \win) p(\stim) d\rate d\stim\\
&= \int \reward \left [ \derivative{}{\win_{ij}} p(\rate|\stim; \win) \right ] p(\stim) d\rate d\stim.
\end{align}

We could theoretically stop here and evaluate $\derivative{}{\win_{ij}} p(\rate|\stim; \win)$ explicitly. However, in the same way that we can approximate $\obj$ as an empirical average over samples, we would like to be able to approximate our derivative as an average. To do this requires us to keep our loss in the form of an expectation over $p(\rate|\stim; \win) p(\stim)$. We notice a convenient identity: $\derivative{}{\win_{ij}} p(\rate|\stim; \win) = \derivative{}{\win_{ij}} \exp (\log p(\rate|\stim; \win)) = \left [\derivative{}{\win_{ij}} \log p(\rate|\stim; \win) \right ] p(\rate|\stim; \win)$, which is a simple application of the chain rule. Inserting this identity into the above equation, we get:

\begin{align} \label{basic_r_update}
\derivative{\obj}{\win_{ij}} &= \int \reward \left [ \derivative{}{\win_{ij}} \log p(\rate|\stim; \win) \right ] p(\rate|\stim; \win) p(\stim) d\rate d\stim \\
&\approx \frac{1}{K} \sum_{k=0}^K R(\rate^{(k)}, \stim^{(k)}) \left [ \derivative{}{\win_{ij}} \log p(\rate^{(k)}|\stim^{(k)}; \win) \right ].
\end{align}

Though this is an approximation, we note that by the Law of Large Numbers, we can improve its accuracy arbitrarily by increasing our number of samples $K$. In practice, however, taking $K=1$ will prove to be the most straightforward way to get an update that is local in time---although such an update will still on average match the true gradient exactly, its high variance can lead to very inefficient learning.

We have left the derivation completely general up until this point. Different choices of $p(\rate|\stim;\weight)$ will produce different updates. Our particular choice gives:

\begin{align}
\derivative{}{\win_{ij}} \log p(\rate|\stim; \win) &= \derivative{}{\win_{ij}} \sum_{i = 0}^{N_r}\frac{1}{2\sigma^2}(\rate_i - f_i(\win \stim))^2 + C \\
&= \frac{1}{\sigma^2}\sum_{n = 0}^{N_r}(\rate_i - f_i(\win \stim))\derivative{f_i(\weight \stim)}{\win_{ij}}.
\end{align}

For a particular weight $ \win_{ij}$, $\derivative{f_l(\win \stim)}{\weight_{ij}} = 0$ if $i \neq l$, so we have:

\begin{align}
 \derivative{}{\win{ij}} \log p(\rate|\stim; \weight) &= \frac{1}{\sigma^2}(\rate_i - f_i(\weight \stim))f^\prime_i(\weight \stim)\stim_j.
\end{align}

Plugging this equation into Eq. \ref{reinforce_update_1} gives the following parameter update:

\begin{equation}
  \Delta \win_{ij} \propto \frac{1}{K} \sum_{k=0}^K R(\rate^{(k)}, \stim^{(k)}) \left [ \frac{1}{\sigma^2}(\rate^{(k)}_i - f_i(\win \stim^{(k)}))f^\prime_i(\win \stim^{(k)})\stim^{(k)}_j \right ] \approx \derivative{\obj}{\win_{ij}}.
\end{equation}

If we want to update all of our parameters simultaneously using parallelized matrix operations, we can write this as an outer product:

\begin{equation}
\Delta \win \propto \frac{1}{K} \sum_{k=0}^K R(\rate^{(k)}, \stim^{(k)}) \left [ \frac{1}{\sigma^2}(\rate^{(k)} - f(\win \stim^{(k)})) \odot f^\prime(\win \stim^{(k)}) \right ] \stim^{(k)T},
\end{equation}
where $\odot$ denotes a Hadamard (elementwise) vector product. Interestingly, the $\frac{1}{\sigma^2}(\rate - f(\win \stim))$ term here is exactly equal to $\noise$.

\subsection{Why don't we need the derivative of the loss?}

One way of interpreting this parameter update is that neural units are correlating fluctuations in their neural activity with the rewards received to approximate $\derivative{\reward}{\rate}$ (Fig. \ref{fig_supp_1}c). To see this, first notice that:

\begin{equation}
\expect{b \left [ \frac{1}{\sigma^2}(\rate - f(\win \stim)) \odot f^\prime(\win \stim) \right ] \stim^T}_{p(\rate|\stim)} = 0,
\end{equation}
for any constant $b$, because $\expect{\rate - f(\win \stim)}_{p(\rate|\stim)} = 0$. If we take $b = \expect{\reward}_{p(\rate|\stim)}$, then we can rewrite the gradient without changing its expected value:

\begin{align}
\derivative{\obj}{\win_{ij}} &= \int (\reward - \expect{\reward}_{p(\rate|\stim)}) \left [ \frac{1}{\sigma^2}(\rate_i - f_i(\win \stim))f^\prime_i(\win \stim)\stim_j \right ] p(\rate|\stim;\win) p(\stim) d\rate d\stim \\
&= \int \frac{1}{\sigma^2} Cov(\reward, \rate_i)\left [f^\prime_i(\win \stim)\stim_j \right ] p(\stim) d\stim,
\end{align}
where $Cov(\reward, \rate_i) = \int (R - \expect{R}_{p(\rate|\stim)}) (\rate_i - \expect{\rate_i}_{p(\rate|\stim)})p(\rate|\stim) d\rate$ is the stimulus-conditioned covariance between network firing rates and reward.
The sample-based parameter update is therefore using the fluctuations in neural activity to compute this covariance.

\subsection{Assessing REINFORCE} \label{assessing_reinforce}
Now that we have derived REINFORCE, we can examine its qualities as a normative plasticity theory. First, we ask: is this algorithm `local' (Section \ref{locality})?
The gradient for a particular synapse, $\derivative{\obj}{\win_{ij}}$ can be approximated with samples in an environment with stimuli $\stim$, firing rates $\rate$, and rewards $\reward$ by $\reward \left [ \frac{1}{\sigma^2}(\rate_i - f_i(\win \stim))f^\prime_i(\win \stim)\stim_j \right ]$. To decide whether this could be a plasticity rule implemented (or more realistically, approximated) by a biological system, we need to think about what pieces of information a synapse would have to have available.

First, the synapse needs $\stim_j$, which amounts to just the presynaptic input, a common feature of any Hebbian synaptic plasticity rule. Second, the synapse needs $\frac{1}{\sigma^2}(\rate_i - f_i(\win \stim))f^\prime_i(\win \stim)$. $\frac{1}{\sigma^2}$ is a constant, and so can be absorbed into the learning rate. $\rate_i$ is the postsynaptic firing rate, which is also a common feature of any Hebbian plasticity rule. $(\win \stim)_i$ is the current injected into the postsynaptic neuron, and $f_i(\cdot)$ and $f^{\prime}_i(\cdot)$ are both monotonic functions of this current, so it is quite conceivable that these values could be approximated by a biochemical process. Third, every synapse needs access to the scalar reward value received on a given trial, $\reward$. This is the most `nonlocal' information involved in the parameter update, however, there exist many theories about how neuromodulatory systems in the brain can deliver information about reward diffusely to many synapses and induce plasticity (Section \ref{locality}).

Now, we have already demonstrated that REINFORCE is able to perform approximate gradient descent for reinforcement learning objective functions---this in itself makes the algorithm very promising as a normative plasticity model (Section \ref{objective}). Its chief advantage is that it does not require detailed knowledge of the reward function $\reward$ (i.e. how to differentiate it), which means that an animal could simply receive a reward from its environment, and relay that reward signal diffusely to its synapses. However, this also restricts the types of objectives that could plausibly be learned by a neural system. Unsupervised learning objectives like the ELBO require detailed knowledge of every neural activity of every neuron in the circuit in order to be calculable (Appendix \ref{wake_sleep_tutorial}), and there is no evidence for downstream neural circuits that perform such calculations. Therefore, even though in principle REINFORCE can be used to train a neural network on \textit{any} objective, explicit reinforcement is much more plausible than other alternatives.

We have only provided a derivation for a single-layer rate-based neural network with additive Gaussian noise, but REINFORCE extends quite readily to multilayer \citep{williams1992simple}, spiking \citep{fremaux2013reinforcement}, and recurrent networks \citep{miconi2017biologically} without any loss of locality. This indicates that the algorithm is both architecture-general (Section \ref{architecture}) and can handle temporal environmental structure (Section \ref{temporal}). Further, because a weight update can be calculated in a single trial, animals could use it to learn online (Section \ref{online}). The biggest point of failure for REINFORCE is that it scales poorly with high complexity in stimuli or task, large numbers of neurons, or prolonged delays in receipt of reward \citep{werfel2003learning,fiete2004learning, bredenberg2021impression}. The greater the number of neurons that contribute to reward and the higher the complexity of the reward function, the harder it becomes to estimate the correlation between a single neuron and reward, which is a prerequisite for the algorithm's function. Thus, though the algorithm is an unbiased estimator of the gradient, it can still be so variable an estimate as to be effectively useless in complex contexts. This suggests that if animals exploit the principles of REINFORCE to update synapses, it is likely an approach paired with other algorithms, or hybridized in a way that allows for better scalability.

The last way to assess REINFORCE is on the basis of how it can be tested (Section \ref{testability}). The simplest way to test this algorithm is by examining whether scalar reward-like signals (i.e. $\reward$) have a multiplicative effect on local plasticity in a circuit. At a single-neuron level this corresponds to identifying neuromodulators that affect plasticity. At a feedback level this corresponds to identifying neuromodulatory systems that project to the circuit in question, and observing whether their stimulation or silencing improves or blocks circuit-level plasticity or behavioral learning performance, respectively. These steps do not identify REINFORCE as the only possibility, but it narrows down the field of possibilities considerably, removing all candidate algorithms that either do not require any feedback, or that require more detailed feedback signals (Fig. \ref{fig_3}a).

\section{Wake-Sleep} \label{wake_sleep_tutorial}
\begin{figure}[t!]
	\includegraphics[width=\linewidth]{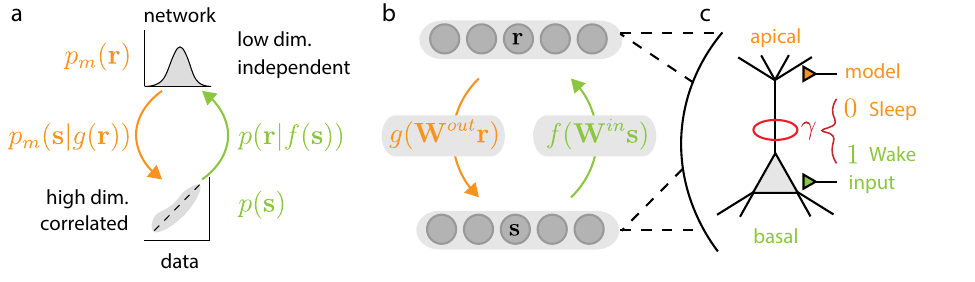}
    \centering
	\caption{\small {\bf The Wake-Sleep algorithm.} {\bf a.} The four components of a good representation: $p_m(\stim|\rate)$ and $p(\rate|\stim)$ map $\rate$ to $\stim$ and back again from $\stim$ to $\rate$, respectively. $p_m(\rate)$ defines `useful' features of a neural representation by constraining its topology. $p(\stim)$ provides the environmental input distribution, which the neural representation must match. {\bf b.} The architecture of the Wake-Sleep algorithm: the decoder, $g(\decoder \rate)$ maps $\rate$ to $\stim$, and the forward map, $f(\win \stim)$ maps $\stim$ to $\rate$. {\bf c.} Physically, these maps correspond to a multicompartmental pyramidal neuron model for each layer, where the `model' synapses are on the apical dendrites, and the `forward map' synapses are on the basal dendrites. $\gamma$ gates which synapses determine neural activity, putting the network in the Wake phase $\gamma = 1$ or the Sleep phase $\gamma = 0$.}
	\label{fig_supp_2}
\end{figure}
Here we will provide a mathematical tutorial on the Wake-Sleep algorithm \citep{hinton1995wake, dayan1995helmholtz}, which is one candidate biologically plausible learning algorithm for constructing a representation in sensory cortices. We will first provide one possible formulation of representation learning as an optimization problem \citep{roweis1999unifying}, and then introduce the Wake-Sleep algorithm\footnote{For another excellent tutorial with more of a machine learning focus, see \citep{kirby2006tutorial}.}, showing how the components necessary to the algorithm could be mapped onto a multicompartmental dendritic neuron model with local synaptic learning. We will then discuss how the algorithm can be extended beyond our simplified introduction.

\subsection{Defining a good objective} \label{ws_objective}

Suppose that at any given moment in time, a neural network is receiving sensory stimuli $\stim$ from its environment. Our first challenge is to articulate what it would mean to form a good neural representation $\rate$ of these stimuli (Fig. \ref{fig_supp_2}a). First of all, `represented' stimuli should be decodable from neural firing rates, i.e. there should exist a mapping $g(\cdot): \mathbb{R}^{N^r} \rightarrow \mathbb{R}^{N^s}$ such that $\stim \approx g(\rate)$. Second, we will also argue that neural firing rates should be decodable from \textit{stimuli}, i.e. there should exist a mapping $f(\cdot): \mathbb{R}^{N^s} \rightarrow \mathbb{R}^{N^r}$ such that $\rate \approx f(\stim)$---this means that there cannot be `extra' features of neural activity that are not contained within the stimuli themselves. This amounts to postulating an approximately bijective relationship between stimuli and firing rates. It means that neural activities should directly correspond to stimuli that have been received.

If these two requirements were sufficient, we might want to simply have one neuron per stimulus dimension, and have it faithfully replicate its immediate input as accurately as possible, i.e. we would take $ f(\stim) = \mathbb{I} \stim$ and $g(\rate) = \mathbb{I} \rate$, where $\mathbb{I}$ is an identity matrix, so that $\rate = \mathbb{I} \stim = \stim$. This identity transformation is obviously not useful, which makes one wonder---what does it mean for a transformation to be useful? Most, if not all unsupervised machine learning and neuroscientific conceptions of a `useful' representation reduce to some formulation of either metabolic or coding efficiency. Approaches within this `efficiency' umbrella include dimensionality reduction \citep{roweis1999unifying}, clustering \citep{illing2021local, dayan1995helmholtz}, gain control \citep{simoncelli1998model}, whitening/factorization \citep{rezende2014stochastic}, and sparsity \citep{simoncelli2001natural}. Each of these definitions of `usefulness' can be formulated as statements about the distribution of neural activities, independent of particular received stimuli, e.g. there are fewer neurons than stimulus dimensions (dimensionality reduction), neural activations occupy roughly discrete clusters in state space (clustering), neurons tend to be uncorrelated with one another (whitening/factorization), or neurons typicaly have low, sparse firing rates (gain control/sparsity/metabolic efficiency). In our formulation, ultimately learning will be unsupervised because we have made \textit{a priori} determinations of what constitutes an efficient representation, and seek to transform incoming data to match those determinations.

Under our definition outlined so far, there are four components of a representation: the stimuli $\stim$ themselves, distributed according to some probability distribution $p(\stim)$ determined by the environment; a decoder, which we will formulate probabilistically as $p_m(\stim|g(\rate; \theta_m))$, which models the probability of $\stim$ given our mapping from neural firing rates $\rate$; a forward mapping from $\stim$ to $\rate$, which we will also formulate probabilistically as $p(\rate|f(\stim; \theta))$; and our definition of efficiency, which dictates how neural firing rates `should' be distributed, independently of stimuli themselves $p_m(\rate)$. Notice that here we have parameterized the forward map $p(\rate | f(\stim; \theta))$ and the decoder (inverse map) $p_m(\stim | g(\rate; \theta_m))$: once we formulate our objective, these will be the parameters that are adjusted to minimize it. $p(\stim)$---the environmental data distribution---obviously cannot change, but we could (and in practice would often want to) parameterize $p_m(\rate)$ and also fit those parameters. We have formulated our four components using probability distributions: after describing our objective function in these terms, we will show one possible way of mapping the components onto neural architecture.

Now, we have evocatively organized our components into two groups: $p_m(\stim | g(\rate; \theta_m))$ and $p_m(\rate)$, versus $p(\stim)$ and $p(\rate | f(\stim; \theta))$. The first group forms a joint distribution $p_m(\rate, \stim; \theta_m)$ which has the subscript $m$ to indicate that it is a generative \textit{model} of the data. Ideally, if its parameters were accurately fit, we could sample $\rate \sim p_m(\rate)$, and then sample $\stim \sim p_m (\stim | g(\rate; \theta_m))$ and get a stimulus that looks like realistic environmental data. The second group also forms a joint distribution $p(\rate, \stim; \theta)$, which amounts to a forward mapping: we could receive a stimulus from the environment, and then have the probability distribution for firing rates $\rate$ that correspond to it. Organizing our models in this way will allow us to achieve biophysical realism: $g(\cdot; \theta)$ and $f(\cdot; \theta)$ will correspond to actual synaptic connections in a model neural network. In practice, ordinary perception as we traditionally conceive of it would correspond to the forward mapping $f(\cdot; \theta)$. Interestingly, at the end of our derivation, it will become clear how an additional representational feature, `detachability' \citep{clark1994doing}---a mechanism to activate neurons in the absence of the sensory stimuli that correspond to them---will be an emergent property of our formulation. We will show how a neural system might be able to leverage the $g(\cdot;\theta_m)$ to accomplish `detachment', which one might imagine mapping perceptually to imagination, planning, prediction, hallucination, or possibly dreaming in different contexts.

For our representation to be good, the forward map should match its inverse, i.e. $p(\rate, \stim; \theta) \approx p_m(\rate, \stim; \theta_m)$. We could imagine formulating many objective functions that could accomplish this goal, but most of them will not accommodate an approximate optimization algorithm that will end up corresponding to a viable normative plasticity model. We will select the Kullback-Liebler (KL) divergence between these two distributions, precisely because it will produce such a normative plasticity model. Notice, though our presentation of the derivation is top-down, it is disingenuous to characterize normative plasticity model development strictly as top-down: locality would not magically emerge from an arbitrary choice of objective function, but rather this choice of objective function is superior to its many alternatives only \textit{because} it produces locality (we won't be able to see why locality emerges until after we have defined $p$ and $p_m$ explicitly and have derived parameter updates). We take our objective function to be:

\begin{align}\label{loss_wake}
    \mathcal{L}_{Wake} &= D_{KL}(p(\rate,\stim; \theta)||p_m(\rate,\stim; \theta_m)) \nonumber \\
    &= \int \ln \left (\frac{p(\rate, \stim; \theta)}{p_m(\rate,\stim; \theta_m)} \right ) p(\rate, \stim; \theta) d\rate d\stim.
\end{align}

We have evocatively named this loss $\mathcal{L}_{Wake}$ because we will be optimizing this objective function during the Wake phase of the algorithm. We will also later appeal to the opposite KL divergence, which we will be optimizing during the Sleep phase:

\begin{align}
    \mathcal{L}_{Sleep} &= D_{KL}(p_m(\rate,\stim; \theta_m)||p(\rate,\stim; \theta)) \nonumber \\
    &= \int \ln \left (\frac{p_m(\rate, \stim; \theta_m)}{p(\rate,\stim; \theta)} \right ) p_m(\rate, \stim; \theta_m) d\rate d\stim.
\end{align}
These objectives share a global minimum ($p_m = p$), if it exists, but are not the same objective function, because unlike a traditional distance metric, the KL divergence is not symmetric. However, \textit{near} the global minimum, they become approximately equivalent \citep{dayan1995helmholtz, bredenberg2021impression}, which will be an important consideration in assessing the convergence properties of the Wake-Sleep algorithm. Unlike REINFORCE, which will work for any reward function $\reward$, the Wake-Sleep algorithm will only work for objectives formulated in this way: in this case the choice of objective function is intimately related to the resultant plasticity rule.

\subsubsection{Equivalence to the Evidence Lower Bound*}
It should be noted that $\mathcal{L}_{Wake}$ has a long history in unsupervised machine learning, and does not always appear in the context of training a sensory representational system through normative plasticity. In fact, minimizing $\mathcal{L}_{Wake}$ is equivalent to minimizing the variational free energy or maximizing the evidence lower bound (ELBO), the objective underlying the variational autoencoder \citep{rezende2014stochastic, kingma2014autoencoding} and the Expectation-Maximization algorithm for latent state models \citep{roweis1999unifying}. Here, to help relate to the broader literature, we will elaborate on this equivalence for the interested reader. This section is a technical aside, which the uninterested reader may safely skip. In traditional machine learning terms, as we will see, the $\mathcal{L}_{Wake}$ objective is equivalent to maximizing the ELBO, and will fit a generative model $p_m(\rate,\stim; \theta_m)$ to data, as well as train a forward map $p(\rate|\stim; \theta)$ to perform approximate Bayesian inference with respect to that model (i.e. we want $p(\rate|\stim; \theta) \approx p_m(\rate|\stim; \theta_m)$). 

To fit a generative model to data, we would typically use maximum likelihood estimation: we would find the parameters of our generative model $p_m(\rate,\stim; \theta_m)$ that match the distribution of data points as accurately as possible by minimizing with respect to $\theta$:

\begin{align} \label{gen_fit}
D_{KL}(p(\stim)||p_m(\stim;\theta_m)) &= \int \ln \left (\frac{p(\stim)}{p_m(\stim; \theta_m)} \right) p(\stim) d\stim.
\end{align}

When this objective is 0, samples drawn from $p_m(\stim; \theta_m)$ will be indistinguishable from samples drawn from $p(\stim)$, indicating that we have an accurate model of the data distribution. But we are not only interested in fitting a generative model: when our network receives a stimulus $\stim$, we would like it to infer the probability distribution over latent representational states that could correspond to that stimulus, $p_m(\rate | \stim; \theta_m)$. However, we haven't defined this quantity, only $p_m(\rate)$ and $p_m(\stim | \rate; \theta_m)$. From a purely machine learning perspective, we might just try to compute $p_m(\rate | \stim; \theta_m)$ explicitly using Bayes' Theorem:

\begin{equation} 
    p_m(\rate | \stim; \theta_m) = \frac{p_m(\rate) p_m(\stim | \rate; \theta)}{\int p_m(\rate) p_m(\stim|\rate; \theta) d\rate d\stim},
\end{equation}
and for simple generative models this might work. However, for complex, nonlinear models, calculating the high-dimensional integral in the denominator analytically is impossible, and approximating it through Monte Carlo methods is time consuming to the point of intractability. This is motivation enough for machine learning applications, but further, it is not clear how biological system could compute such an integral rapidly upon receiving a single stimulus. So instead, we might try a different approach. We can take our explicitly defined and parameterized forward map $p(\rate | \stim; \theta)$ and train it to approximate $p_m(\rate | \stim; \theta_m)$ as closely as possible by minimizing the expected KL divergence:

\begin{align} \label{inf_fit}
    \expect{D_{KL}(p(\rate | \stim; \theta) || p_m(\rate | \stim; \theta_m))}_{p(\stim)} &= \int \ln \left (\frac{p(\rate|\stim; \theta)}{p_m(\rate | \stim; \theta_m)} \right )p(\rate|\stim; \theta) p(\stim) d\rate d\stim.
\end{align}
If objective is approximately 0, then we do not need to perform Bayes' theorem to calculate the posterior $p_m(\rate|\stim;\theta_m)$, because we have access to a perfect (or near-perfect) approximation $p(\rate|\stim;\theta)$ that we can calculate explicitly or sample from. If $p(\rate|\stim;\theta)$ is parameterized appropriately, this is usually much easier, and potentially could be implemented by a neural network. Now we have two objectives that we want to minimize: one to fit our generative model, and the other to perform approximate inference. It seems natural to add them and minimize them jointly. First, we notice that adding our second objective defines the following inequality:

\begin{align}
    D_{KL}(p(\stim)||p_m(\stim;\theta_m)) &\leq D_{KL}(p(\stim)||p_m(\stim;\theta_m)) + \expect{D_{KL}(p(\rate | \stim; \theta) || p_m(\rate | \stim; \theta_m))}_{p(\stim)}.
\end{align}
due to the positivity of the KL divergence. Second, we note that adding these two objectives together really just gives us $\mathcal{L}_{Wake}$:

\begin{align}
 D_{KL}(p(\stim)||p_m(\stim;\theta_m)) + \expect{D_{KL}(p(\rate | \stim; \theta) || p_m(\rate | \stim; \theta_m))}_{p(\stim)} &= D_{KL}(p(\rate, \stim; \theta) || p_m(\rate, \stim; \theta)) \\
 &= \mathcal{L}_{Wake},
 \end{align}
 where the first equality follows from adding Eqs. \ref{gen_fit} and \ref{inf_fit} and using the properties of the logarithm and expectations.
 
 This alternative construction demonstrates that minimizing that our objective function $\mathcal{L}_{Wake}$ trains our system to perform two separate model-fitting functions: training a generative model and training an approximate inference distribution. From here we can also see its equivalence to the variational free energy and the ELBO:
 
\begin{align}
D_{KL}(p(\stim)||p_m(\stim;\theta_m)) & \leq \mathcal{L}_{Wake} \\
& = \int \ln \left (\frac{p(\rate,\stim; \theta)}{p_m(\rate, \stim; \theta_m)} \right ) p(\rate,\stim; \theta)d\rate d\stim\\
& = \int \ln \left (\frac{p(\rate|\stim; \theta)}{p_m(\rate, \stim; \theta_m)} \right ) p(\rate,\stim; \theta)d\rate d\stim + \int \left (\ln p(\stim) \right ) p(\rate|\stim; \theta) p(\stim)d\rate d\stim\\
&= \int \ln \left (\frac{p(\rate|\stim; \theta)}{p_m(\rate, \stim; \theta_m)} \right ) p(\rate,\stim; \theta)d\rate d\stim + \int \left ( \ln p(\stim) \right ) p(\stim)d\stim.
\end{align}
Now, by definition $D_{KL}(p(\stim)||p_m(\stim;\theta_m)) = \int \left ( \ln p(\stim) \right ) p(\stim)d\stim - \int \left ( \ln p_m(\stim;\theta_m) \right ) p(\stim)d\stim$, the first term of which also appears on the right hand side of our inequality. Furthermore, $\int \left ( \ln p(\stim) \right ) p(\stim)d\stim$ is not a function $\theta_m$ or $\theta$, so from the perspective of optimization, it is an irrelevant additive constant. We subtract it from both sides to get:
\begin{align}
    - \int \left ( \ln p_m(\stim;\theta_m) \right ) p(\stim)d\stim \leq \int \ln \left (\frac{p(\rate|\stim; \theta)}{p_m(\rate, \stim; \theta_m)} \right ) p(\rate,\stim; \theta)d\rate d\stim.
\end{align}
This expression on the left is the negative log-likelihood, and the expression on the right is the variational free energy, which is the negative of the ELBO. This shows that $\mathcal{L}_{Wake}$ and the variational free energy differ only by an additive constant from the perspective of optimization: minimizing one is the same as minimizing the other. Similarly, $\mathcal{L}_{Sleep}$ corresponds to an upper bound on the reverse KL divergence, $D_{KL}(p_m(\stim; \theta_m) || p(\stim))$.

\subsection{Defining $p$ and $p_m$}
Let us start by selecting three features of our representation that we think will be useful, i.e. efficient. First, we want our neurons to be metabolically efficient: a biological system cannot have neurons wasting energetic resources by firing too much \citep{simoncelli2003vision}. One way of requiring this would be to stipulate that the squared norm of our neural firing rate vector, $\|\rate \|_2^2$ lies within some reasonable range of activation values. Second, we want to reduce the dimensionality of our representation: many naturalistic datasets are low-dimensional, and it may be wasteful to represent some high-dimensional features of stimuli that are just due to sensor noise. To accomplish this, we will stipulate that $N_r \ll N_s$, where $N_r$ is the representation's dimensionality, and $N_s$ is the stimulus dimension. Third, we will require that individual neural activations should be independent from one another, which will allow individual neurons to extract important features of the data without requiring full knowledge of the activity of other neurons in the representation. To achieve a representation that embodies these three desired features, we define $p_m(\rate)$ as follows:
\begin{equation}
    p_m(\rate) \sim \mathcal{N}(0,1),
\end{equation}
i.e. we will require that the representation, averaged over stimuli, will match an $N_r$-dimensional multivariate normal distribution, where individual axes $\rate_i$ are independent from one another (uncorrelated), and where the normal distribution naturally restricts the probable range of neural activities to lie within bounds determined by the variance (arbitrarily set to 1). Though this distribution captures several intuitions for how neural representations should function, it is clearly a toy model for several reasons: it does not restrict firing rates to be positive, it does not allow for activities to be discrete spikes, it does not account for temporal dynamics, etc. We will discuss later how each of these extensions have been done before, but for now, many features of our model $p_m(\rate, \stim)$ and our forward map $p(\rate | \stim)$ will be unrealistic for didactic purposes.

Now we define the probabilistic decoder $p_m(\stim | \rate)$ (Fig. \ref{fig_supp_2}b), which takes neural firing rates and produces estimates of stimuli, as follows:
\begin{equation}
    p_m(\stim | \rate; \decoder) \sim \mathcal{N}(g(\decoder \rate), \sigma_s),
\end{equation}
where $g(\cdot)$ is an arbitrary nonlinearity, and $\sigma_s^2$ is the variance of the decoder. In this probability distribution, we will treat the $N_s \times N_r$ matrix $\decoder$ as a free parameter which we will train to optimize our objective.

Similarly, we can define the forward map $p(\rate | \stim)$, which takes environmental stimuli and produces firing rates, as follows:
\begin{equation}
    p(\rate|\stim; \win) = \mathcal{N}(f(\win \stim), \sigma_r),
\end{equation}
where $f(\cdot)$ is an arbitrary (potentially different) nonlinearity, and $\sigma_r$ will ultimately correspond to intrinsic neural variability. Here, the $N_r \times N_s$ matrix $\win$ is the free parameter. Thus, $\win$ and $\decoder$, are the free parameters in our simple construction.

We have not yet made clear how these parameters and functions could map onto an actual neural architecture: we will do this after defining the learning algorithm, so that it is clear what the necessary components of the algorithm are. Interestingly, we do not have to define $p(\stim)$ at all. This distribution is determined by the environment. In fact, a learning system should ideally be as agnostic as possible to the specific form of $p(\stim)$ as possible, in order to be able to adapt strange and unforeseen changes in the statistics of the world. The Wake-Sleep algorithm is ideal in that it makes little-to-no assumption about $p(\stim)$, but as we will see, it may perform poorly if it is not possible to obtain a close match between $p$ and $p_m$. This might occur if the environmental distribution of $\stim$ is much higher dimensional than the number of neurons, or is in some other way more complex than the generative model.
\subsection{Approximating the loss gradient}
Having defined our objective function and probability distributions $p$ and $p_m$, we can now derive the Wake-Sleep algorithm. First, we will show that we can obtain a promising update for $\decoder$ by performing gradient descent on $\mathcal{L}_{Wake}$ (the Wake phase of learning). We will next show that we can obtain a similarly promising update for $\win$ by performing gradient descent on $\mathcal{L}_{Sleep}$ (the Sleep phase of learning). One might easily wonder why we did not perform gradient descent on $\mathcal{L}_{Wake}$ with respect to $\win$, instead of $\mathcal{L}_{Sleep}$: we will next show why it would be a bad idea to do this. Lastly, we will describe two perspectives on how these resultant updates can be viewed as a unified form of approximate optimization.

\subsubsection{Wake}
We start by calculating the negative gradient of $\mathcal{L}_{Wake}$ with respect to a particular parameter $\decoder_{ij}$ from $p_m(\stim | \rate; \decoder)$:

\begin{align} \label{eq_wake_1}
 -\derivative{\mathcal{L}_{Wake}}{\decoder_{ij}} &= - \derivative{}{\decoder_{ij}} \int \ln \left (\frac{p(\rate, \stim; \win)}{p_m(\rate,\stim; \decoder)} \right ) p(\rate, \stim; \win) d\rate d\stim \\
 &= - \derivative{}{\decoder_{ij}} \int \left [\ln p(\rate, \stim; \win) - \ln p_m(\stim | \rate; \decoder) - \ln p_m(\rate) \right ]p(\rate, \stim; \win) d\rate d\stim \\
 &= - \int \derivative{}{\decoder_{ij}} \left [\ln p(\rate, \stim; \win) - \ln p_m(\stim | \rate; \decoder) - \ln p_m(\rate) \right ]p(\rate, \stim; \win) d\rate d\stim \\
 &= \int \left [ \derivative{}{\decoder_{ij}} \ln p_m(\stim | \rate; \decoder) \right ]p(\rate, \stim; \win) d\rate d\stim
\end{align}

Plugging in the probability density function for $p_m(\stim | \rate; \decoder) \sim \mathcal{N}(g(\decoder \rate), \sigma^2_s)$, we end up with:

\begin{equation}\label{eq_wake_2}
     -\derivative{\mathcal{L}_{Wake}}{\decoder_{ij}} = \int \left [ \derivative{}{\decoder_{ij}} \frac{1}{2\sigma_s^2} \sum_{i=0}^{N_s}(\stim - g(\decoder \rate))^2 \right ]p(\rate, \stim; \win) d\rate d\stim.
\end{equation}

Similar to our derivation for REINFORCE, we see that for a particular weight $ \decoder_{ij}$, $\derivative{g_l(\decoder \rate)}{\decoder_{ij}} = 0$ if $i \neq l$. Thus, we have:

\begin{align}
    -\derivative{\mathcal{L}_{Wake}}{\decoder_{ij}} & = \int \frac{1}{\sigma_s^2}\left [ (\stim_i - g_i(\decoder \rate)) g'_i(\decoder \rate) \rate_j \right ]p(\rate, \stim; \win) d\rate d\stim.
\end{align}

Again, similar to REINFORCE, we can approximate this update as the network actively `perceives': we receive a sampled environmental stimulus $\stim^{(k)}$, and then sample from the probability distribution $p(\rate | \stim^{(k)}; \win)$ to obtain a firing rate sample $\rate^{(k)}$. Then across $K$ samples, we calculate the approximate parameter update:

\begin{align}  \label{wake_update}
     \Delta \decoder_{ij} \propto \frac{1}{\sigma_s^2 K} \sum_{k=0}^K \left [ (\stim^{(k)}_i - g_i(\decoder \rate^{(k)})) g'_i(\decoder \rate^{(k)}) \rate^{(k)}_j \right ] \approx -\derivative{\mathcal{L}_{Wake}}{\decoder_{ij}}.
\end{align}

If we want learning to be able to occur online (Section \ref{online}), then we can take $K = 1$, and sacrifice some precision of our estimate. This update has the form of a prediction error, where the the error between the true stimulus $\stim_i^{(k)}$ and the network's decoded estimate $g_i(\decoder \rate^{(k)})$ combine with presynaptic inputs $r_j^{(k)}$ to produce parameter updates. In Section \ref{assessing_ws} we will analyze in detail how this parameter update could correspond to a local synaptic update for a particular neuron model.

\subsubsection{Sleep}
So far, other than performing stochastic gradient descent over $K$ samples, we have introduced no approximation into our algorithm. We might be tempted to perform gradient descent on $\mathcal{L}_{Wake}$ with respect to $\win$ too: though we will defer the discussion of this point for later, it turns out to be a bad idea (see Section \ref{ws_not_reinforce}). Instead, we will perform an \textit{almost identical} procedure, but perform gradient descent on $\mathcal{L}_{Sleep}$ instead. As discussed in Section \ref{objective}, one way of interpreting this change in loss is that we now have two different sets of parameters (i.e. synapses) in our system, $\win$ and $\decoder$ which are optimizing two different, albeit closely related objectives, $\mathcal{L}_{Sleep}$ and $\mathcal{L}_{Wake}$, respectively. An alternative perspective that we will discuss is that $\win$ is also optimizing $\mathcal{L}_{Wake}$, but is only performing an approximate gradient descent. We will discuss in Section \ref{convergence} how this added complexity affects the convergence and quality of the algorithm. Starting with $\mathcal{L}_{Sleep}$, we have:

\begin{align}
 -\derivative{\mathcal{L}_{Sleep}}{\win_{ij}} &= - \derivative{}{\win_{ij}} \int \ln \left (\frac{p_m(\rate, \stim; \decoder)}{p(\rate,\stim; \win)} \right ) p_m(\rate, \stim; \decoder) d\rate d\stim \\
 &= \int \left [ \derivative{}{\win_{ij}} \frac{1}{2\sigma_r^2} \sum_{i=0}^{N_r}(\rate - f(\win \stim))^2 \right ]p_m(\rate, \stim; \decoder) d\rate d\stim,
\end{align}
where we have followed exactly the same steps as in Eqs. \ref{eq_wake_1}-\ref{eq_wake_2}.

As before, we notice that for a particular weight $ \win_{ij}$, $\derivative{f_l(\win \stim)}{\win_{ij}} = 0$ if $i \neq l$. Thus, we have:

\begin{align}
    -\derivative{\mathcal{L}_{Sleep}}{\win_{ij}} & = \int \frac{1}{\sigma_r^2}\left [ (\rate_i - f_i(\win \stim)) f'_i(\win \stim) \stim_j \right ]p_m(\rate, \stim; \decoder) d\rate d\stim.
\end{align}

Now we can approximate this update with samples from $p_m(\rate, \stim; \decoder)$. Notice that we are no longer actively perceiving via the forward mapping $p(\rate|\stim)$ in response to sampled environmental stimuli. Instead, activity is first internally generated via $\rate^{(k)} \sim p_m(\rate)$, before propagating to the stimulus layer to produce artificial stimuli via $\stim^{(k)} \sim p_m(\stim | \rate^{(k)}; \decoder)$. This is termed the Sleep phase of the algorithm evocatively: an animal could not perform this type of learning while actively moving through an environment, and if it did perceive, such percepts would appear hallucinatory or dream-like, being reflective of the animal's model rather than reality. Given our $K$ samples, we calculate the approximate parameter update:

\begin{align} \label{sleep_update}
      \Delta \win_{ij} \propto \frac{1}{\sigma_r^2 K} \sum_{k=0}^K \left [ (\rate^{(k)}_i - f_i(\win \stim^{(k)})) f'_i(\win \stim^{(k)}) \stim^{(k)}_j \right ] \approx -\derivative{\mathcal{L}_{Sleep}}{\win_{ij}}.
\end{align}

Now, this update should look almost equivalent to the Wake update for $\decoder$ (Eq. \ref{wake_update}). As with the Wake update, if we want learning to occur online we can take $K = 1$. It turns out that the variability induced by this sampled approximation is \textit{much} less than the variability induced by the REINFORCE algorithm, and is the chief reason for its superior performance and scalability \citep{bredenberg2021impression}. However, it is very important to note that we are sampling from $p_m$ instead of $p$. Because our two parameter updates, Eq. \ref{wake_update} and Eq. \ref{sleep_update} require sampling from two different probability distributions and individual neurons $\rate$ could only be sampling from one probability distribution at a time, the updates are necessarily computed during different \textit{phases}. The Wake-Sleep algorithm consists of alternating between sampling from $p$ to compute updates for $\decoder$ (the Wake phase; Eq. \ref{wake_update}) and sampling from $p_m$ to compute updates for $\win$ (the Sleep phase; Eq. \ref{sleep_update}). As we discuss in Section \ref{assessing_ws}, we should be appropriately cautious about what these alternative phases could possibly mean for a biological organism.

\subsection{Assessing Wake Sleep}\label{assessing_ws}
Having derived our Wake-Sleep parameter updates, we are finally in a position to assess the degree to which it satisfies our desiderata. We have provided a very simplified derivation of the Wake-Sleep algorithm, for a single-layer rate-based network. However, the algorithm generalizes well to recurrent, spiking, and multilayer architectures \citep{dayan1996varieties} (Section \ref{architecture}), and these modifications do make the algorithm more realistic as a normative plasticity model. However, it will still be very useful to show how the various components of the algorithm as we have derived it could potentially map onto realistic biological structures (Fig. \ref{fig_supp_2}c). First of all, we observe that both $\stim$ and $\rate$ need to be able to sample from either $p_m$ or $p$---for this to be possible, $\stim$ must be \textit{internal} to the brain, since sampling from $p_m$ affects both $\rate$ and $\stim$ simultaneously and would have to occur while an animal is not consciously acting in its environment. Therefore, it is best to think of $\stim$ as a stimulus layer of neurons, and of $\rate$ as a downstream layer of neurons receiving feedforward inputs. Next, we suppose that there is a global gating signal $\gamma$ that determines the phase of the network--- if $\gamma = 1$, the network is in the Wake phase, and if $\gamma = 0$, the network is in the Sleep phase. Now we observe that the following equations will produce valid samples:

\begin{align}
    \rate &= \gamma f(\win \stim) + (\gamma \sigma_r + (1-\gamma)) \noise_r \\
    \stim &= \gamma \stim_{p} + (1-\gamma) (g(\decoder \rate) + \sigma_s \noise_s),
\end{align}
where $\stim_p \sim p(\stim)$ is an incoming sensory input, and $\noise_s, \noise_r \sim \mathcal{N}(0,1)$ are sources of intrinsic noise for neurons in the stimulus, and downstream layers, respectively. Because $p_m$ and $p$ both assume exactly the same dimensionality of $\rate$ (and $\stim$), the only reasonable mapping of these two different sampling phases is onto one neuron with two different \textit{modes} of activity. In Figure \ref{fig_supp_2}c, we show that one possible biological mapping is to propose that feedforward inputs (active when $\gamma = 1$) to the basal dendrites of pyramidal neurons allow neurons to sample from $p$, and top-down inputs (active when $\gamma = 0$) to the apical dendrites of pyramidal neurons allows neurons to sample from $p_m$: interestingly, a corollary of this mapping is that a network could achieve `detachability' by manipulating $\gamma$ to generate sample network states in the absence of stimuli.

It is important to note that several normative plasticity models have proposed that top-down signals to the apical dendrites could serve as some form of training signal. We will adopt a similar attitude, and now assess the locality of the Wake-Sleep parameter updates with respect to this model formulation. If we take the sample size for our updates to be $K=1$, based on Eqs. \ref{wake_update} and \ref{sleep_update}, for a single pair of samples $\rate$, $\stim$, we have:

\begin{align} \label{sampling_eqs1}
    \Delta \win_{ij} \propto \frac{1-\gamma}{\sigma_r^2} \left [ (\rate_i - f_i(\win \stim)) f'_i(\win \stim) \stim_j \right ] \\
    \Delta \decoder_{ij} \propto \frac{\gamma}{\sigma_s^2} \left [ (\stim_i - g_i(\decoder \rate)) g'_i(\decoder \rate) \rate_j \right ]. \label{sampling_eqs2}
\end{align}

As with REINFORCE, both $\sigma_r$ and $\sigma_s$ are proportionality constants and can be disregarded. For $\Delta \win_{ij}$, a basal synapse on $\rate_i$, several variables are required. First, the same signal that gates the influence of apical versus basal inputs, $\gamma$, must also \textit{deactivate} plasticity at basal synapses. $\gamma$ could be implemented in a neural circuit by either global inhibitory gating or by a neuromodulatory signal \citep{bredenberg2021impression}---whichever candidate signal would also have to gate plasticity. The synapse needs the postsynaptic firing rate $\rate_i$, which is readily available, and a subtracted measure of current local to the basal compartment, $f_i(\win \stim)$---there is some indication that local dendritic voltage levels can affect synaptic plasticity, but the sign and exact form of this effect is variable across studies \citep{letzkus2006learning, froemke2005spike, sjostrom2006cooperative}. As with REINFORCE, the synapse would require $f'_i(\win \stim)$, which is simply a monotonic function of $(\win \stim)_i$, and could be easily approximated; lastly, it would need the presynaptic firing rate $\stim_j$. The information requirements for $\decoder_{ij}$ are almost exactly the same. 

In terms of requiring only functions of pre- and postsynaptic activity, with the addition of some limited global context signal $\gamma$, these plasticity rules are plausibly local (Section \ref{locality}). However, several features of this setup are unconfirmed, the most obviously testable being the Wake-Sleep sampling dynamics postulated by Eqs. \ref{sampling_eqs1} and \ref{sampling_eqs2}: it seems unlikely that a neural network would entirely and synchronously switch into a `generative' or hallucinatory regime for an extended period of time when $\gamma = 0$, and such a regime could not possibly occur in an awake, behaving animal, meaning that $\win$ could not be learned online (Section \ref{online}). However, a softer form of Wake-Sleep has been proposed \citep{bredenberg2021impression} which does allow for online learning, and does not interfere with active perception, suggesting that the principles established by Wake-Sleep may extend to more realistic formulations of $\gamma$. The strongest test (Section \ref{testability}) of this family of algorithms is that artificially magnifying the influence of apical dendrites in a neural circuit should induce generative sampling, i.e. hallucination; other models of apical dendritic learning \citep{sacramento2017dendritic, guerguiev2017towards, payeur2021burst, urbanczik2014learning} do not propose this as a mechanism. Notice that this prediction requires our specific mapping of the Wake-Sleep algorithm onto neural circuitry: other interpretations are conceivable, and would have different predictions.

As we have discussed in Section \ref{ws_objective}, the Wake-Sleep algorithm is capable of optimizing a broad range of \textit{unsupervised} learning objectives, considerably more general than for instance Oja's rule \citep{oja1982simplified} (though the specific toy example we provide is just a nonlinear form of probabilistic PCA). Unlike REINFORCE, the Wake-Sleep algorithm is unable to optimize reinforcement learning objectives, however, within the range of objectives that Wake-Sleep \textit{can} optimize, it is typically much more scalable than REINFORCE (Section \ref{scalable})\footnote{Though it still performs worse than backpropagation \citep{kingma2014autoencoding, rezende2014stochastic}.}: in this way, it is an ideal complement, and having both algorithms or some hybridized form present in a neural circuit could be very powerful. However, the Wake-Sleep algorithm involves more approximation than REINFORCE. One could very easily wonder: since we have presented two sets of parameters in the Wake-Sleep algorithm minimizing two different objective functions, why should we expect the algorithm to converge or reliably improve performance on either objective? 

To this point, we have identified two strange features of the Wake-Sleep algorithm that go hand-in-hand. First, it is strange that we should require a period of hallucinatory activity to train our parameters. Second, it is hard to interpret the convergence of an algorithm that is alternatively minimizing two slightly different objective functions: why all the work and extra conceptual baggage? Why not just do approximate gradient descent as we did with the REINFORCE algorithm and be done with it?  In Section \ref{ws_not_reinforce} we will motivate why more standard gradient descent methods are not appropriate for this type of unsupervised learning, and in Section \ref{convergence} we will address the convergence properties of the Wake-Sleep algorithm from two different perspectives, explaining why the algorithm has such good empirical performance despite its approximations.

\subsubsection{Why gradient descent with $\win$ won't work}\label{ws_not_reinforce}
Sometimes, to genuinely understand an algorithm, it's important to understand the weaknesses of alternative approaches. For didactic reasons, we will explore what happens if we simply take the gradient of $\mathcal{L}_{Wake}$ with respect to $\win$. We have:

\begin{align}
 -\derivative{\mathcal{L}_{Wake}}{\win_{ij}} = &- \derivative{}{\win_{ij}} \int \ln \left (\frac{p(\rate, \stim; \win)}{p_m(\rate,\stim; \decoder)} \right ) p(\rate, \stim; \win) d\rate d\stim \\
 = &- \int \left [ \derivative{}{\win_{ij}} \ln \left (\frac{p(\rate, \stim; \win)}{p_m(\rate,\stim; \decoder)} \right ) \right ] p(\rate, \stim; \win) d\rate d\stim \nonumber\\
 &- \int \ln \left (\frac{p(\rate, \stim; \win)}{p_m(\rate,\stim; \decoder)} \right ) \derivative{}{\win_{ij}} p(\rate, \stim; \win) d\rate d\stim \\
 = &- \int \left [ \derivative{}{\win_{ij}} \ln \left (p(\rate, \stim; \win) \right ) \right ] p(\rate, \stim; \win) d\rate d\stim \nonumber\\
 &- \int \ln \left (\frac{p(\rate, \stim; \win)}{p_m(\rate,\stim; \decoder)} \right ) \derivative{}{\win_{ij}} p(\rate, \stim; \win) d\rate d\stim, \label{ws_not_reinforce_eq_1}
\end{align}

where the second equality follows from the product rule, and the third equality follows from the fact that $\ln p_m (\rate, \stim; \decoder)$ does not depend on $\win$. Interestingly, the first term in this equation is zero. To see this, we note the following sequence of identities:

\begin{align}
    \int \left [ \derivative{}{\win_{ij}} \ln \left (p(\rate, \stim; \win) \right ) \right ] p(\rate, \stim; \win) d\rate d\stim &= \int \left [ \derivative{}{\win_{ij}} e^{\ln \left (p(\rate, \stim; \win) \right )} \right ] d\rate d\stim \nonumber\\
    &= \int \derivative{}{\win_{ij}} p(\rate, \stim; \win) d\rate d\stim \nonumber\\
    &= \derivative{}{\win_{ij}} \int p(\rate, \stim; \win) d\rate d\stim = \derivative{}{\win_{ij}} 1 = 0 \label{log_derivative}.
\end{align}

The first term is zero, which leaves only the second term of Eq. \ref{ws_not_reinforce_eq_1}. It gives us:

\begin{align}
     -\derivative{\mathcal{L}_{Wake}}{\win_{ij}} &= - \int \ln \left (\frac{p(\rate, \stim; \win)}{p_m(\rate,\stim; \decoder)} \right ) \derivative{}{\win_{ij}} p(\rate, \stim; \win) d\rate d\stim \\
     &= \int \ln \left (\frac{p_m(\rate,\stim; \decoder)}{p(\rate, \stim; \win)} \right ) \left (\derivative{}{\win_{ij}} \ln p(\rate, \stim; \win) \right ) p(\rate, \stim; \win) d\rate d\stim, \label{reinforce_like_ws_update}
\end{align}

where for the second equality we have once again used the identity in Eq. \ref{log_derivative}. Fascinatingly enough, this is exactly equivalent to the REINFORCE update (Eq. \ref{basic_r_update}), if we take $R(\rate, \stim) = \ln \left (p/p_m \right )$. Though the REINFORCE update might be practical for environmental rewards that an animal might receive, this particular choice of $R(\rate, \stim)$ requires detailed knowledge of the inner workings of a neural representation. Not only is it not possible for an environmental signal to carry this information, there is no evidence that any neuromodulatory center in the brain is able to compute such a complicated signal based on neural network activity. Thus, even though this update appears to have the form of a reward-modulated Hebbian plasticity rule, there is very little reason to believe that it is local (Section \ref{locality}). Furthermore, this form of update is well-known to have severe scalability (Section \ref{scalable}) issues, and demonstrably performs worse than Wake-Sleep on high-dimensional datasets \citep{werfel2003learning, bredenberg2021impression}. The Wake-Sleep algorithm is very much a response to these failings, using a local error signal specific to each neuron, rather than correlating each neuron's activity with a global reward signal. However, the Wake-Sleep algorithm employs more approximations than REINFORCE. In Section \ref{convergence}, we will analyze the convergence properties of Wake-Sleep.

\subsubsection{The convergence of Wake-Sleep}\label{convergence}
Currently, we have two updates that are approximating gradient descent on two different objectives: $\Delta \decoder_{ij} \approx - \lambda \derivative{\mathcal{L}_{Wake}}{\decoder_{ij}}$, and $\Delta \win_{ij} \approx - \lambda \derivative{\mathcal{L}_{Sleep}}{\win_{ij}}$, where $\lambda$ is a small positive learning rate. In Section \ref{objective}, we stressed the importance of viewing plasticity updates as decreasing a \textit{unified} objective, but here we have two. How do we know that $\Delta \win_{ij}$ won't \textit{increase} $\mathcal{L}_{Wake}$ and vice versa? Clearly, $\mathcal{L}_{Sleep}$ and $\mathcal{L}_{Wake}$ are closely related: one way of resolving this difficulty is by demonstrating that $\Delta \win_{ij} \approx - \lambda \derivative{\mathcal{L}_{Wake}}{\win_{ij}}$. In this case, during the Wake phase, the system would optimize $\mathcal{L}_{Wake}$ with respect to $\decoder$, and during the Sleep phase, it would approximately optimize the same objective with respect to $\win$---this would amount to an approximation of coordinate descent. In fact, under certain conditions, it turns out that this is exactly what the Wake-Sleep algorithm is doing. 

To see this, we begin with the REINFORCE-like update (Eq. \ref{reinforce_like_ws_update}) for gradient descent on $\mathcal{L}_{Wake}$:

\begin{align}
    -\derivative{\mathcal{L}_{Wake}}{\win_{ij}}
     &= \int \ln \left (\frac{p_m(\rate,\stim; \decoder)}{p(\rate, \stim; \win)} \right ) \left (\derivative{}{\win_{ij}} \ln p(\rate, \stim; \win) \right ) p(\rate, \stim; \win) d\rate d\stim.
\end{align}

Interestingly, we notice that if $p_m \approx p$, then by first-order Taylor expansion, $\ln \left (p_m/p \right ) \approx p_m/p - 1$. Plugging this approximation in (see \citep{bredenberg2021impression} for a more detailed justification of this approximation), we get:

\begin{align}
    -\derivative{\mathcal{L}_{Wake}}{\win_{ij}}
     &\approx \int \left (\frac{p_m(\rate,\stim; \decoder)}{p(\rate, \stim; \win)}  - 1\right ) \left (\derivative{}{\win_{ij}} \ln p(\rate, \stim; \win) \right ) p(\rate, \stim; \win) d\rate d\stim \\
     &= \int \left (\derivative{}{\win_{ij}} \ln p(\rate, \stim; \win) \right ) p_m(\rate, \stim; \decoder) d\rate d\stim \\
     &= - \derivative{\mathcal{L}_{Sleep}}{\win_{ij}},
\end{align}

where for the first equality we have once again used the identity Eq. \ref{log_derivative}. Essentially, if a global optimum such that $p_m = p$ exists, it is shared by both $\mathcal{L}_{Wake}$ and $\mathcal{L}_{Sleep}$. Thus, we can expect the gradients of these two objective functions to behave very similarly if $p_m$ is close to $p$. Because the Wake phase (updating $\decoder$) occurs without approximation, the algorithm has the opportunity to enter this regime before the approximating Sleep phase ever occurs.

An alternative analysis of the Wake-Sleep algorithm \citep{dayan1995helmholtz} observes that for fixed $\decoder$, $\mathcal{L}_{Sleep}$ and $\mathcal{L}_{Wake}$ share a global minimum with respect to $\win$ when $p_m(\rate | \stim; \decoder) = p(\rate | \stim; \win)$, as long as there exists a $\win_{opt}$ such that this equality holds. If $\mathcal{L}_{Sleep}$ is convex and this global minimum is attainable, fully optimizing $\mathcal{L}_{Sleep}$ with respect to $\win$ during the Sleep phase is therefore guaranteed to also optimize $\mathcal{L}_{Wake}$. Therefore, as long as these to conditions of convexity and attainability of the global minimum are satisfied (they are not in general, but do hold for simple examples like Factor Analysis \citep{amari1999convergence}), both phases decrease $\mathcal{L}_{Wake}$. Rather than an approximation of coordinate descent, this can be viewed as an approximation of the Expectation-Maximization (EM) algorithm \citep{dempster1977maximum}.

We see that there are two different ways of interpreting Wake-Sleep: first, it is an approximation of coordinate descent that becomes a better approximation the closer to the optimum it becomes. Second, under restricted conditions, Wake-Sleep can be viewed as an approximation of the EM algorithm. Both of these perspectives are conditional on assumptions about the probability models being trained, requiring a generative model $p_m(\rate, \stim)$ and a forward map $p(\rate|\stim)$ capable of mutually reaching good performance for an environmental stimulus distribution $p(\stim)$. Though Wake-Sleep empirically performs quite well under a variety of stimulus conditions and network models \citep{dayan1996varieties}, these are important caveats: the comparative weakness of the demonstrations of Wake-Sleep's convergence relative to gradient descent or EM is a common point of criticism of the algorithm \citep{rezende2014stochastic, kingma2014autoencoding, mnih2014neural}.

\end{appendix}
\end{document}